\documentclass[prl,nofootinbib,superscriptaddress,twocolumn]{revtex4-2}
\pdfoutput=1
\usepackage[utf8]{inputenc}

\usepackage[english]{babel}

\usepackage{tikz}
\usetikzlibrary{shapes,backgrounds,fit,decorations.pathreplacing,arrows,decorations.markings}
\usepackage{verbatim}
\tikzstyle{vecArrow} = [thick, decoration={markings,mark=at position
   1 with {\arrow[semithick]{open triangle 60}}},
   double distance=1.4pt, shorten >= 5.5pt,
   preaction = {decorate},
   postaction = {draw,line width=1.4pt, white,shorten >= 4.5pt}]
\tikzstyle{innerWhite} = [semithick, white,line width=1.4pt, shorten >= 4.5pt]
\usepackage{bbm}
\usepackage{bm}
\usepackage{amsbsy}
\usepackage{amsthm}
\usepackage{amssymb}
\usepackage{amsfonts}
\usepackage{amsmath}
\usepackage{mathtools}
\usepackage{dsfont} 
\usepackage{graphicx} 
\usepackage{epsfig}
\usepackage{epstopdf}
\usepackage{dsfont}
\usepackage{color}

\usepackage[colorlinks]{hyperref}
\makeatletter
\newcommand\org@hypertarget{}
\let\org@hypertarget\hypertarget
\renewcommand\hypertarget[2]{%
  \Hy@raisedlink{\org@hypertarget{#1}{}}#2%
  }
\makeatother
\usepackage[figure,table]{hypcap}
\usepackage{MnSymbol}
\usepackage{enumerate}
\usepackage
{enumitem}
\usepackage{float}
\hypersetup{
	bookmarksnumbered,
	pdfstartview={FitH},
	citecolor={darkgreen},
	linkcolor={darkred},
	urlcolor={darkblue},
	pdfpagemode={UseOutlines}}
\definecolor{darkgreen}{RGB}{50,190,50}
\definecolor{darkblue}{RGB}{0,0,190}
\definecolor{darkred}{RGB}{238,0,0}
\definecolor{quantum}{RGB}{83,37,127}
\definecolor{quantumlight}{RGB}{169,146,191}
\usepackage{soul}
\newcommand{\mpel}{}

\newcommand{\ket}[1]{\ensuremath{\left|\right.\!{#1}\!\left.\right\rangle}}

\newcommand{\bra}[1]{\ensuremath{\left\langle\right.\!{#1}\!\left.\right|}}
\newcommand{\braket}[2]{\ensuremath{\langle{#1}|{#2}\rangle}}
\newcommand{\ketbra}[2]{\ensuremath{|{#1}\rangle\!\langle{#2}|}}

\newcommand{\tr}{\textnormal{Tr}}

\makeatletter
\renewcommand{\p@subsection}{}
\renewcommand{\p@subsubsection}{}

\makeatother

\newcommand{\PRLapp}[1]{Appendix~\ref{#1}}
\newcommand{\sys}{\mathrm{S}}

\begin{document}

\title{Quantum measurements and equilibration:\\the emergence of objective outcomes via entropy maximisation}

\author{Emanuel Schwarzhans}
\email{emanuel.schwarzhans@oeaw.ac.at} 
\affiliation{Atominstitut, Technische Universit{\"a}t Wien, 1020 Vienna, Austria}
\affiliation{University of Malta}

\author{Felix C. Binder}
\email{quantum@felix-binder.net} 
\affiliation{School of Physics, Trinity College Dublin, Dublin 2, Ireland}

\author{Marcus Huber}
\email{marcus.huber@tuwien.ac.at} 
\affiliation{Atominstitut, Technische Universit{\"a}t Wien, 1020 Vienna, Austria}
\affiliation{Institute for Quantum Optics and Quantum Information - IQOQI Vienna, Austrian Academy of Sciences, Boltzmanngasse 3, 1090 Vienna, Austria}

\author{Maximilian P.~E. Lock}
\email{maximilian.paul.lock@tuwien.ac.at} 
\affiliation{Atominstitut, Technische Universit{\"a}t Wien, 1020 Vienna, Austria}
\affiliation{Institute for Quantum Optics and Quantum Information - IQOQI Vienna, Austrian Academy of Sciences, Boltzmanngasse 3, 1090 Vienna, Austria}

\date{\today}

\begin{abstract}
The measurement postulate of quantum theory stands in conflict with the laws of thermodynamics and has evoked debate regarding what actually constitutes a measurement. With the help of modern quantum statistical mechanics, we take the first step in formalising the hypothesis that quantum measurements are driven by the natural tendency of closed systems to maximize entropy, a notion that we call the Measurement-Equilibration Hypothesis. In this paradigm, we investigate how classical measurement outcomes can emerge within a purely unitary framework, and find that: (i) the interactions used in standard measurement models fail to spontaneously encode information classically and (ii) while ideal projective measurements are impossible, one can (for a given form of Hamiltonian) approximate them exponentially well as more physical systems are collected together into an ``observer'' system. We thus lay the groundwork for self-contained models of quantum measurement, proposing improvements to our simple scheme.
\end{abstract}

\maketitle

\raggedbottom

\section{Introduction} 

The mathematical formulation of quantum theory includes two types of dynamics for a closed quantum system: unitary evolution, and measurement. The latter is often referred to as a collapse of the wavefunction, instantaneously updating the state of the system, and is at the heart of many interpretational discussions of quantum mechanics (e.g. the ``measurement problem''). For projective measurements, this is an apparent contradiction of the laws of thermodynamics, as it does not conserve energy, allows for an arbitrary decrease in entropy~\cite{Guryanova_2020}, and allows for cooling to zero-temperature with finite resources, in violation of Nernst's formulation of the $3^\text{rd}$ law \cite{nernst1906, taranto_2021}. In this work, we investigate a potential resolution to \mpel{the apparent tension between thermodynamics and quantum measurement} by modelling the latter as the system of interest and its environment undergoing a closed-system equilibration process, thus being driven by an \emph{increase} in entropy.

In the widely-used von Neumann measurement scheme~\cite{von1930mathematische}, subsequently developed into what is sometimes referred to as the standard model of measurement~\cite{Busch_1996}, a carefully-timed interaction correlates the system of interest with some probe or pointer system. This allows, for example, an investigation of the restrictions placed by quantum theory on incompatible observables. Quantum Darwinism~\cite{zurek2003decoherence,zurek2006decoherence,zurek2009quantum} broadens this model to include the environment, and thus the observer itself, highlighting the role of the redundant encoding of information about the system state in the measurement basis (the ``pointer basis'') into the environment degrees of freedom, arguably removing the need for an explicit ``Heisenberg cut''. Demanding that this redundancy be such that multiple observers (i.e. parts of the environment) independently
agree on the system state and that the only correlations between them be their shared information about the system, implies that the total state of the system and observers necessarily exhibits Spectrum Broadcast Structure~\cite{Horodecki2015,Le2019,korbicz2021roads}. The appearance of this structure has been demonstrated theoretically for a dielectric sphere in a photonic environment~\cite{korbicz2014objectivity}, in a spin-spin model~\cite{Mironowicz2017,Mironowicz2018}, and in a Quantum Brownian Motion model~\cite{tuziemski2016}.

The process of measurement, specifically the transformation from quantum indeterminacy to an objective fact, is irreversible and occurs without precise control of all microscopic degrees of freedom of the system's environment. A dynamical model of measurement must therefore explain how the system and its environment evolve spontaneously and irreversibly towards a state structure exhibiting objectivity. \mpel{In the following, we explore the hypothesis that, rather than being necessarily in competition with the second law of thermodynamics, the spontaneous, irreversible transition from quantum to classical is in fact an entropy-increasing transition towards an equilibrium state, and refer to this as the \textit{Measurement-Equilibration Hypothesis}}.



The notion that measurements correspond to an increase in entropy has been discussed at various points since the early years of quantum theory, \textit{e.g.}~\cite{szilard1929entropieverminderung,von1930mathematische,daneri1962quantum,misra1979lyapounov,peres1980can,zurek1986maxwell,allahverdyan2013understanding,nussinov2020a,nussinov2022exact}, but is absent from dynamical models of measurement. Recent advances in quantum statistical mechanics, specifically in the equilibration of closed quantum systems~\cite{gogolin2016equilibration} allow us to address this absence, and as a first step in this direction, we seek to determine the circumstances under which a general system and environment equilibrate to states exhibiting classical properties, including the Spectrum Broadcast Structure mentioned above.  To avoid potential confusion, we note that in the present context, we consider equilibration in the general sense, \textit{i.e.} spontaneous evolution towards some equilibrium state, rather than the specific example of thermalization, whereby the equilibrium state is a Gibbs state.


\mpel{We begin by describing in detail the physical basis of the hypothesis which we introduce here in Sec.~\ref{sec:Motivation}, before reviewing some elements of the theory of equilibration of closed quantum systems in Sec.~\ref{sec:EqOnAv}, emphasizing how instantaneous properties at equilibrium are determined by the time-averaged state. Before investigating whether this theory can describe the quantum-to-classical transition, one must first define what is meant by classical, and we therefore discuss how quantum states can encode classical information, as well as the concepts of objectivity and Spectrum Broadcast Structure, in Sec.~\ref{sec:States}. We ask how the equilibration process constrains the dynamics capable of producing states with a classical structure in Sec.~\ref{sClassPropsEq}, and in Sec.~\ref{sExactImpossible} use this to show that it is impossible to reach states with this structure exactly. In Sec.~\ref{sStanMod} we show how the standard model of measurement does not result in an equilibrium state exhibiting any correlations whatsoever between the system and its environment, unless one explicitly includes the free evolution (in contrast with the model's usual assumption). We then show in Sec.~\ref{sec:ApproximateSBS} how a classical state structure, and in fact the stricter condition of objectivity, is approximated exponentially well with increasing size of the ``observers'' for whom the measurement outcome should be objective.}
Thus, the equilibration process can (approximately) copy a microscopic observable into multiple macroscopic ones. We finish with a discussion of the consequences of our findings and the problems raised by our analysis, as well as future directions and open questions.


\section{\protect\mpel{Statistical mechanics and quantum measurements}} \label{sec:Motivation}

\mpel{Understanding the emergence of irreversible behavior in isolated systems is a central goal of statistical mechanics. In classical theory, the approach to equilibrium in isolated systems is explained through statistical arguments illustrating the typical properties of dynamical trajectories of many-body systems~\cite{khinchin1949,penrose1979foundations}. In the quantum domain, it has become clear in recent decades how equilibration and thermalization phenomena arise in isolated many-body systems~\cite{gogolin2016equilibration}. Equilibration occurs due to the tendency of certain observables to quickly approach, and remain indistinguishably close to, their time-averaged value during most of the system's evolution~\cite{reimann2008foundation,short2011equilibration}, a process sometimes referred to as ``equilibration on average''. Thermalization occurs when the energy eigenstates constituting this time-averaged (i.e.~equilibrium) state have certain ``typical'' properties~\cite{dalessio2016quantum}. Such emergent irreversible behavior occurs only for highly-degenerate observables of many-body systems~\cite{Anza2018}, for example those local to subsystems, or those describing bulk properties. This highlights the sense in which such equilibration is a form of the second law of thermodynamics, arising from the many-to-one relationship between the microscopic states of the system and the outcomes of equilibrating observables~\cite{meier2024emergence}. In an interacting chain of many spin systems, for example, such observables include quantities associated with a single spin, or the bulk magnetization of the entire chain.}

\mpel{The theory of Quantum Darwinism highlights the fact that the transition from quantum to classical entails redundantly copying information from the system of interest in the complement of that system~\cite{zurek2003decoherence}, which one may interpret as a passive environment or as being associated with some observer systems. The encoding of information from a smaller system into a larger one necessarily implies the degeneracy of this encoding, and as the size mismatch between systems increases, the dynamics driving such an encoding will (unless fine-tuned) generically lead to the effective irreversibility described above. In the context of the Schr\"{o}dinger's cat thought experiment: there are a vast number quantum states of the cat compatible with the decayed atom (i.e.\ death) and with the undecayed atom (i.e.\ life), and so the dynamics coupling the atomic decay and the cat is necessarily irreversible with respect to this encoding.\footnote{Here we use ``cat'' as a shorthand for its combination with some minimal environment required for its continued, short-term survival in the case of the undecayed atom.}}

\mpel{Here we adopt the hypothesis that this emergent irreversibility is a defining feature of the measurement process, and thus that the second law of thermodynamics drives the quantum-to-classical transition. We then seek to explain phenomena such as the stable appearance of a classical plateau over time, as in e.g.~\cite{doucet2024classifying}, in an analogous manner to the tendency of particle speeds in an ideal gas to stably tend to the Maxwell-Boltzmann distribution, despite the ongoing microscopically reversible evolution of the system in both cases. }

\mpel{However, an important distinction must be made: states exhibiting Quantum Darwinism or objectivity are generally incompatible with thermality~\cite{riedel2012rise,le2021thermality}. In particular, the thermal state represented by the Gibbs distribution, effectively erases all information about the system’s initial conditions except that encoded by the temperature. Clearly, not all entropy-increasing processes qualify as measurements; the latter demands that information concerning the observed quantity be preserved. This does not pose a problem for the hypothesis explored here, since not all equilibration processes correspond to thermalization. Observables pertaining to systems with many conserved quantities may still equilibrate~\cite{rigol2007relaxation}, approaching a stable distribution over time that does not correspond to a thermal state. The equilibrium state is given by maximizing entropy while preserving with all constants of motion in the initial state~\cite{gogolin2016equilibration}, and thus may retain much more information about the initial state than in the case of thermalization.}

\mpel{To assess the viability of the Measurement-Equilibration hypothesis, one must therefore investigate to what extent states exhibiting features such as Quantum Darwinism are compatible with the theory of equilibration of isolated systems, in particular outside of the case of thermalization. As we shall see, the framework of equilibration on average permits a continued microscopic coherent evolution in tandem with a macroscopic behavior that becomes effectively stationary and classical in appearance. In the following section we illustrate the sense in which such distinct microscopic and macroscopic descriptions can coexist.}

\section{Equilibration of isolated quantum systems} \label{sec:EqOnAv}

\begin{figure}[ht!]
\begin{center}
\includegraphics[width=0.8\columnwidth]{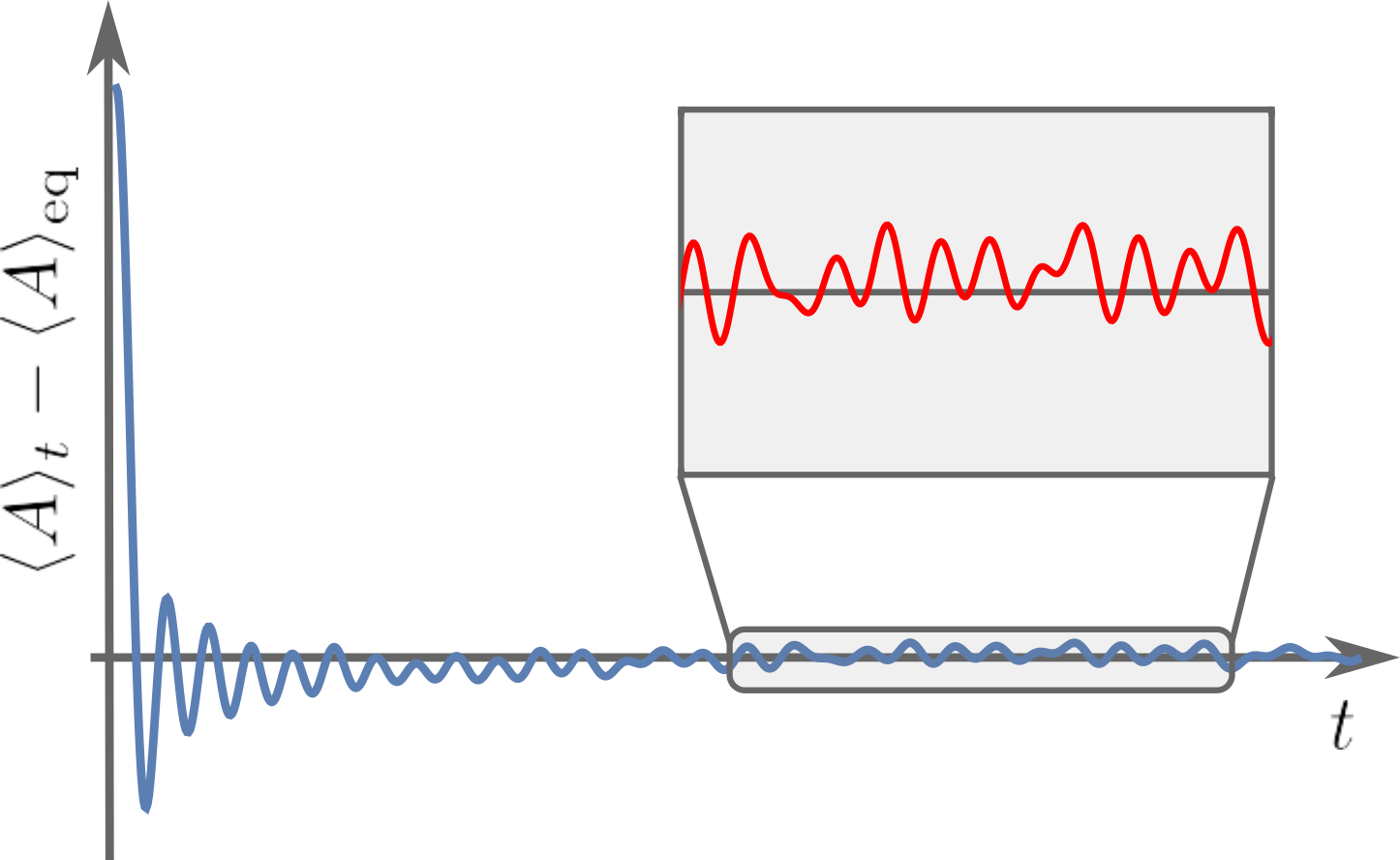}
\caption{Illustration of the equilibration (on average) of an observable $A$ over time.  The difference between the instantaneous expectation value of an observable $A$ at some time $t$ and its equilibrium value, decreases over time, despite the system continuing to evolve unitarily. The inset illustrates finite fluctuations around the equilibrium, which generally depend on the effective dimension of the system, in accordance with Eq.~\eqref{Eq:reimannbound}.}
\label{fig:pendula}
\end{center}
\end{figure}

To investigate how objective measurement outcomes emerge as a result of equilibration, we will employ the notion of equilibration on average~\cite{gogolin2016equilibration}, whereby the equilibrium state is given by the infinite-time average. That is, if a system evolving as $\rho(t)$ equilibrates on average, then the equilibrium state is given by
\begin{align} \label{eRhoEqDefn}
     \rho_\mathrm{eq} := \lim_{T\rightarrow \infty}\dfrac{1}{T}\int_0^T dt \, \rho(t)  .
 \end{align}
This is the unique state which maximises the von Neumann entropy given all conserved quantities~
\cite{gogolin2011absence,gogolin2016equilibration}. To illustrate the sense in which states equilibrate to their time average, we now review two examples of equilibration on average, consider the following example. Let the evolution be generated by some time-independent Hamiltonian $H$, so that~$\rho(t)=U(t)\rho(0)U^\dag(t)$ with $U(t)=e^{-i H t}$, and for simplicity assume that no two energy gaps of $H$ are equal in magnitude (which also implies that $H$ is non-degenerate). Denoting by $\lbrace \rho_{nn}\rbrace_{n}$ the diagonal elements of $\rho(0)$ in the eigenbasis of $H$, the effective dimension (sometimes called the inverse participation ratio) is defined by $d_\mathrm{eff}:=(\sum_{n}\rho_{nn}^{2})^{-1}$. Note that this quantity is time-independent. Now, starting from the initial state $\rho(0)$ and selecting a future state $\rho(t)$ randomly according to a uniform distribution across $t\in[0,\infty]$, the expectation value $\tr [A \rho(t)]$ of a bounded Hermitian operator $A$ satisfies~\cite{reimann2008foundation}
\begin{equation}\label{Eq:reimannbound}
    \mathrm{Prob} \left[ \Big\vert \tr [A \rho(t)] - \tr [A \rho_\mathrm{eq}]  \Big\vert \geq \frac{\Delta_{A}}{d_\mathrm{eff}^{1/3}}\right] <\frac{1}{d_\mathrm{eff}^{1/3}} 
\end{equation}
with $\Delta_{A}:=a_\mathrm{max}-a_\mathrm{min}$, where $a_\mathrm{max}$ and $a_\mathrm{min}$ are respectively the highest and lowest values of $A$ that have a non-zero probability according to $\rho(0)$. Thus, regardless of the initial value $\tr [A \rho(0)]$, when the effective dimension is high, and when $\Delta_{A}<<d_\mathrm{eff}^{1/3}$, then we will observe statistics close to the equilibrium value of $A$ for most of its future trajectory through the Hilbert space (as illustrated in Fig.~\ref{fig:pendula}). Indeed, averaging over the entire future trajectory of the state, we have~\cite{short2011equilibration}
\begin{equation}
    \left\langle \Big\vert \tr [A\rho(t)]-\tr [A\rho_\mathrm{eq}] \Big\vert^{2} \right\rangle \leq \frac{\Delta_{A}^{2}}{4 d_\mathrm{eff}} .
\end{equation}
We stress that equilibration on average does not mean that $\rho (0)$ unitarily evolves to $\rho_\mathrm{eq}$, but rather that the statistics of $A$ for the unitarily-evolving state $\rho(t)$ will be close to those of $\rho_\mathrm{eq}$ for most times, as illustrated in Fig.~\ref{fig:pendula}.

As a second example, consider $\rho(t)$ to instead be some subsystem of a larger, unitarily evolving system, \textit{i.e.}~$\rho(t)=\tr_{X}[U(t)\tilde{\rho}(0)U^\dag(t)]$ for some $\tilde{\rho}(0)$ and tensor factor $X$, and again assume that no two energy gaps of $H$ are equal. In this case, the trace distance $D$ between $\rho(t)$ and $\tr_{X}[\tilde{\rho}_\mathrm{eq}]$ averaged over the entire future trajectory of the state satisfies~\cite{linden2009quantum}:
\begin{equation}
    \langle D\left[\rho(t),\tr_{X}[\tilde{\rho}_\mathrm{eq}]\right] \rangle \leq \frac{1}{2}\sqrt{\frac{d^{2}}{d_\mathrm{eff}}} ,
\end{equation}
where $d$ is the dimensionality of the Hilbert space to which the subsystem $\rho(t)$ belongs, and $d_\mathrm{eff}$ is calculated with respect to the total system $\tilde{\rho}(0)$.

We note three issues for both of these examples to be considered equilibration. First, both rely on an effective dimension $d_\mathrm{eff}$ being in some sense large. This quantity measures the number of significantly occupied energy levels participating in the evolution of $\rho(t)$, and its typically-large size for large systems can be argued for on both physical and mathematical grounds. Second, there is the necessity for some finite equilibration time, and third, the possibility of Poincar\'{e} recurrences. All three of these issues, as well as generalisations beyond the condition of no equal energy gaps, are addressed in detail in~\cite{gogolin2016equilibration}. Furthermore, we stress that equilibration does not in general imply thermalisation; we discuss the relevance of this distinction to the present work in Sec.~\ref{sec:Discussion}.

In the following, we will assume that $d_\mathrm{eff}$ is large enough that one can speak of the isolated system equilibrating on average, with negligible fluctuations. It will thus suffice to examine the structure of the equilibrium state given in Eq.~\eqref{eRhoEqDefn}, as any observable that equilibrates on average, equilibrates to this state~\cite{gogolin2016equilibration}.

\section{Measurement and quantum states} \label{sec:States}

Let $\mathcal{H}_\sys$ denote the Hilbert space corresponding to some $d_\sys$-dimensional system of interest, and let $\lbrace\mathcal{H}_{k}\rbrace$ with $k=1,\ldots,N$ denote the Hilbert spaces corresponding to $N$ systems which form the environment of the system of interest, and which we will associate with observers. Note that we do not \emph{a priory} impose any criteria for what is to be considered as an observer. To measure the system of interest $\sys$ in some pointer basis, denoted $\lbrace \ketbra{i}{i} \rbrace$, where $i=1,\ldots,d_\sys$, the observer systems must interact with it. One may then collect desiderata for the post-interaction state on $\mathcal{H}_\mathrm{S}\bigotimes_{k=1}^N\mathcal{H}_k$, and use them to constrain its form.

For example, the state which generically represents a lack of coherence in the pointer basis, and the measurement outcomes being encoded in the state of the environment, the ``classical-quantum'' state~\cite{devetak2004distilling}, has the form:
\begin{equation}    
\rho_{\mathrm CG}:=\sum_{i=1}^{d_\sys} p_i \, \ketbra{i}{i}_\sys \otimes \rho_{E}^{(i)},
\label{eq:classical-quantum}
\end{equation}
where the $\lbrace\rho_{E}^{(i)}\rbrace_{i}$ are some density matrices on $\mathcal{H}_{E}:=\bigotimes_{k=1}^{N}\mathcal{H}_{k}$. These states are closely related to Holevo's bound, which sets an upper limit $\chi$ on the amount of classical information that can be encoded in a quantum system using an ensemble $\{p_i, \rho_i\}$~\cite{Nielsen2000}. For classical-quantum states as in Eq.~\eqref{eq:classical-quantum}, the Holevo Information $\chi$ of the environment equals the quantum mutual information $S(S:E)$ between system and environment.

Additionally, in order for a fact about the system to be considered objective (and thus classical), multiple observers must be able to discover it in such a way that they agree with each other and that none of them changes the fact through their observation. As noted above, formalising this notion in the context of quantum theory~\cite{ollivier2004objective} and combining it with the requirement that observers be uncorrelated except via the system (``strong independence''), is equivalent to demanding that the state of the system and the observers exhibit Spectrum Broadcast Structure~\cite{Horodecki2015}. Associating each environmental subsystem $k$ with an observer, then a state has Spectrum Broadcast Structure with respect to a pointer basis $\lbrace \ketbra{i}{i} \rbrace$, where $i=1,\ldots,d_\sys$, if it is of the form
\begin{align} \label{eSBSdefn}
    \rho_\mathrm{SBS}=\sum_{i=1}^{d_\sys} p_i \, \ketbra{i}{i}_\sys \bigotimes_{k=1}^{N} \rho_k^{(i)},
\end{align}
where $p_i$ is the probability associated with outcome $\ket{i}$, and $\rho_k^{(i)}$ is the state on $\mathcal{H}_{k}$ conditioned on that outcome, and where $\rho_k^{(i)}\rho_k^{(j)}=0$ for $i\neq j$. \mpel{Thus, if the dynamics coupling system and environment lead to a state close to the one in Eq.~\eqref{eSBSdefn}, then there exist observables on each part of the  environment which will exhibit the same statistics $\lbrace p_i\rbrace$ as the measured observable on $\sys$, and comparing the outcomes of such observables associated with different parts of the environment will result in agreement~\cite{korbicz2021roads}.} As in~\cite{tuziemski2016}, we will find it more convenient to express the orthogonality of conditional states in terms of the fidelity, namely $F(\rho_k^{(i)},\rho_k^{(j)})=0$ for $i\neq j$, where $F(\sigma,\rho):=\tr\left(\sqrt{\sqrt{\sigma}\rho\sqrt{\sigma}} \right)^2$.

The structure in Eq.~\eqref{eSBSdefn} thus represents a state where quantum information has become redundantly encoded in each observer system, and through this can be considered classical~\cite{blume2006quantum} (though this is, of course, not the only possible definition of classicality). Note, however, that this is only true in the pointer basis, in accordance with the no-cloning theorem~\cite{Wootters1982}. This state structure holds if and only if the conditions of strong Quantum Darwinism and Strong Independence are met~\cite{Le2019}. A projective measurement can then be understood as a post-selection performed on the process by which an initially uncorrelated state between system and observers tends to a state with Spectrum Broadcast Structure in the measurement basis. Consequently, to say that a projective measurement has been performed, one must first arrive at a state structure exhibiting objectivity.

\section{Encoding measurement outcomes via closed-system equilibration} \label{sClassPropsEq}

With this framework, we can now invoke the Measurement-Equilibration Hypothesis, and ask if states such as those described in Sec.~\ref{sec:States} can emerge as a consequence of equilibration, and thus the maximisation of entropy. Specifically, we ask whether there exist conditions under which equilibration can lead to macroscopic observables encoding the outcomes of microscopic observables, and the extent to which multiple macroscopic observables can simultaneously encode a microscopic observable.

\begin{figure}[]
\begin{center}
\includegraphics[width=0.8\columnwidth]{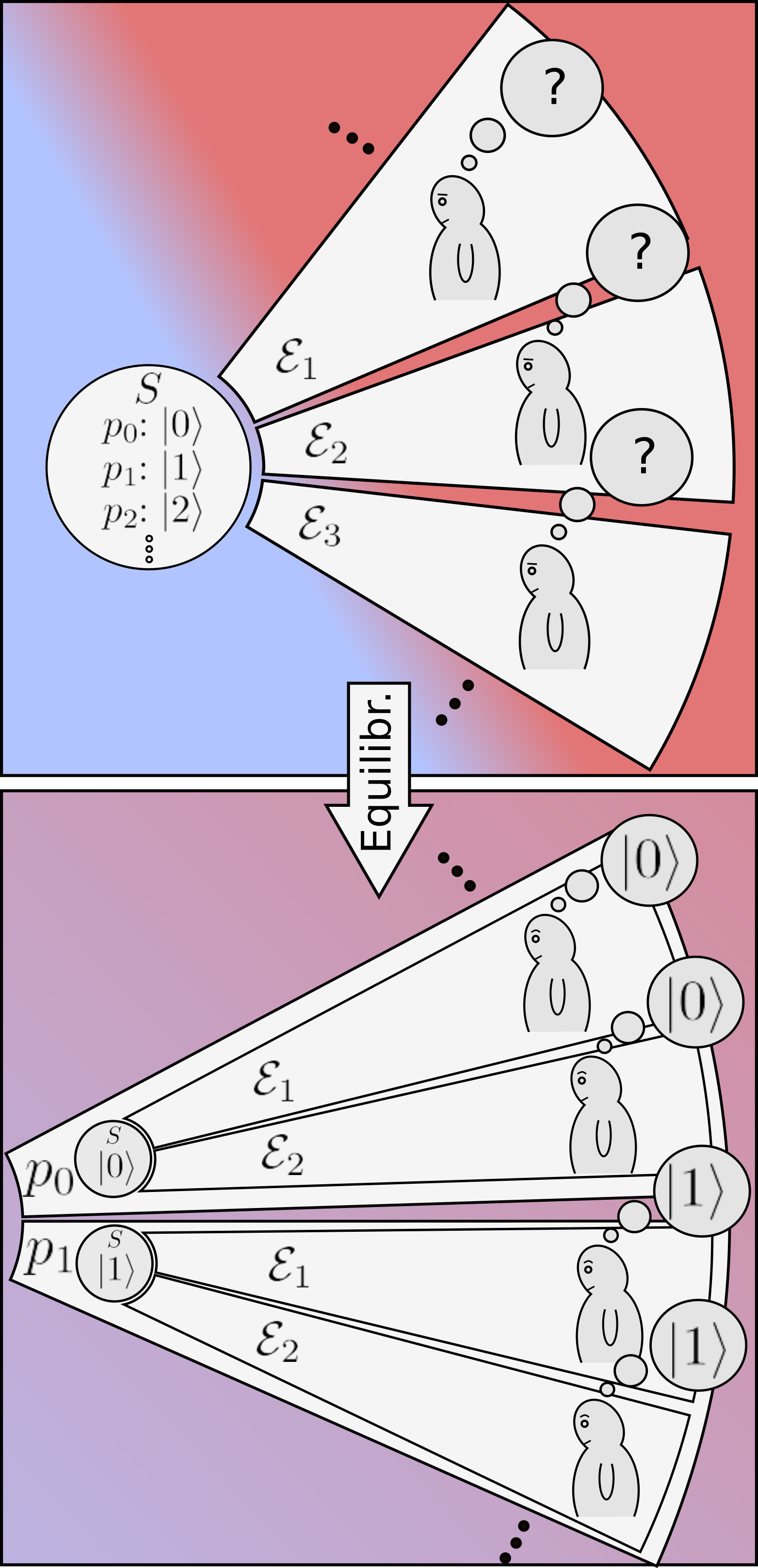}
\caption{Measurement as an equilibration process. The system of interest $S$ and some observer systems $\lbrace\mathcal{E}_{q}\rbrace_{q}$ begin in an uncorrelated, out-of-equilibrium initial state. During the measurement, the collective system undergoes closed-system equilibration, correlating them in the measurement basis $\lbrace\ket{i}\rbrace_{i}$, and encoding the corresponding probabilities $\lbrace p_i=\bra{i}\rho_{\sys ,0}\ket{i}\rbrace_{i}$ in the observer systems in a redundant and objective (i.e. classical) manner.}
\label{fig:EquilibrationToObjectivity}
\end{center}
\end{figure}

To wit, we assume that system and observers are initially uncorrelated and investigate the possible structures of equilibrium states between system and environment, or potential observer systems. We will find the circumstances under which the equilibrium state has classical-quantum form $\rho_\mathrm{eq}=\rho_\mathrm{CQ}$, the extent to which environment states can encode measurement outcomes, and the extent to which it is possible to have $\rho_\mathrm{eq}=\rho_\mathrm{SBS}$. Note that the latter demand does not restrict the observer states $\rho_{k}^{(i)}$ in Eq.~\eqref{eSBSdefn}, beyond the orthogonality requirement described above. As above, we denote the measurement basis on $\mathcal{H}_\sys$ by $\lbrace\ket{i}\rbrace_{i}$, and the outcome probabilities by $\lbrace p_i\rbrace_{i}$, i.e.~$p_i=\bra{i}\rho_{\sys ,0}\ket{i}$, where $\rho_{\sys ,0}$ is the initial state of the system. The scenario is illustrated in Fig.~\ref{fig:EquilibrationToObjectivity}

The infinite-time average is a pinching map with respect to the Hamiltonian generating the evolution~\cite{gogolin2011absence}, \textit{i.e.}
\begin{equation} \label{ePinch}
    \rho_\mathrm{eq} = \sum_{E} \Pi_{E} \rho(0) \Pi_{E} , 
\end{equation}
where $ H=\sum_{E}E \,\Pi_{E}$ is the spectral decomposition of $ H$, and we therefore have $[ H,\rho_\mathrm{eq}]=0$. Consequently, to equilibrate to a state with a given structure, the Hamiltonian must commute with all states of that structure. In particular, for classical-quantum states, we must have $[ H,\rho_\mathrm{CQ}]=0$, which can be understood as an extension of the commutation relation defining the pointer basis~\cite{zurek1981pointer} to include the observers' systems. As we discuss below, no Hamiltonian exists such that any environmental system in $\rho_\mathrm{CQ}$ perfectly encodes the measurement outcome, which precludes equilibration to a state commensurate with an exact projective measurement.

\section{Exact projective measurements are impossible} \label{sExactImpossible}

Consider the initial state ${\rho(0)=\rho_{\sys,0}\otimes\rho_{E,0}}$, where $\rho_{E,0}$ is an arbitrary state on $\mathcal{H}_{E}:=\bigotimes_{k=1}^N\mathcal{H}_{k}$ (and may therefore violate the strong independence condition~\cite{Horodecki2015}). Demanding that the equilibrium state $\rho_\mathrm{eq}$ has the form of a classical-quantum state $\rho_{\mathrm CG}$ defined in Eq.~\eqref{eq:classical-quantum} implies that $ H$ must take the form
\begin{equation} \label{eConditionalForm}
 H=\sum_{i} \ketbra{i}{i}_\sys\otimes H_\mathrm{E}^{(i)} ,    
\end{equation}
for some Hermitian matrices $\lbrace H_{E}^{(i)}\rbrace_{i}$ on $\bigotimes_{k=1}^{N}\mathcal{H}_{k}$. Note that Eq.~\eqref{eConditionalForm} can only be considered as a sum of a free system Hamiltonian $ H_\sys$ and some interaction $ H_{int}$ in the case where the pointer basis consists of eigenstates of $ H_\sys$. For measurement in any other basis, it corresponds to an interaction term alone, necessitating the so-called measurement limit, often used in dynamical models of measurement, where free terms in the Hamiltonian are neglected \textit{i.e.}~$ H\approx H_{int}$ (see \textit{e.g.}~\cite{Busch_1996}). One can understand this as a manifestation of the Wigner-Araki-Yanase theorem~\cite{wigner1952messung,araki1960measurement}, which implies that observables that do not commute with the free Hamiltonian cannot be measured exactly. In this non-commuting case, Eq.~\eqref{eConditionalForm} can hold approximately if the interaction is sufficiently strong that the free evolution can be neglected in the course of equilibration, as we discuss in Sec.~\ref{sec:Discussion}.

When we now demand, in addition to the above, that the environment unambiguously encodes the system state, \textit{i.e.} that $F(\rho_E^{(i)},\rho_E^{(j)})=0$ for $i\neq j$, we find that this only holds when $\rho_{E}^{(i)}=0$ $\forall i$  (see \PRLapp{appsec:constraining} for proof). In particular, $F(\rho_k^{(i)},\rho_k^{(j)})$ is bounded below by $\tr(\rho_{E,0}^2)/d_{E}^{2}$, where $d_{E}$ is the dimensionality of $\bigotimes_{k=1}^{N}\mathcal{H}_{k}$. Thus, it is impossible to perfectly encode a microscopic observable into a macroscopic one, purely through the process of equilibration. 

The impossibility of perfect encoding in the environment additionally precludes equilibration to a state exhibiting objectivity, as defined in~\cite{Horodecki2015}. The stronger condition of Spectrum Broadcast Structure (which combines objectivity with Strong Independence) is then likewise impossible. In~\cite{Guryanova_2020}, the authors found that it was possible to perform an ideal projective measurement given access to a zero-temperature pointer state, but here we see that, adopting the Measurement-Equilibration Hypothesis, it is impossible even given access to such a resource. That the lower bound of $F(\rho_k^{(i)},\rho_k^{(j)})$ rapidly decreases with dimensionality
however, does allow the approximation an objective state as the size of the environment increases, as we will discuss in Section~\ref{sec:ApproximateSBS}.

\section{Equilibration and the standard model of measurement} \label{sStanMod}

As mentioned above, dynamical models of measurement usually assume a degree of temporal control such that a particular unitary transformation can be performed on the system and some apparatus/environment, generated the product of observables acting on the system and the apparatus~\cite{Busch_1996}. For example, one such model was used to investigate the transition to Spectrum Broadcast Structure in~\cite{tuziemski2016}. In our context, let us denote such a Hamiltonian by $H=X_\sys\otimes Y_\mathrm{E}$, where $X_\sys$ is any operator whose eigenbasis is the pointer basis, and where $Y_\mathrm{E}$ is an observable on $\bigotimes_{k=1}^{N}\mathcal{H}_{k}$. Note that, writing $X_\sys$ in terms of its spectral decomposition, this Hamiltonian can be rewritten in the form of Eq.~\eqref{eConditionalForm}, where the $\lbrace H_{E}^{(i)}\rbrace_{i}$ differ from each other by only a scalar factor. In this case, as shown in \PRLapp{appsec: SBS-unattainable with common Hamiltonian}, the equilibrium state is given by
\begin{equation}
    \rho_\mathrm{eq} = \sum_{i=1}^{d_\sys} p_i \, \ketbra{i}{i}_\sys \otimes \rho_E ,
\end{equation}
for some state $\rho_E$ on $\bigotimes_{k=1}^{N}\mathcal{H}_{k}$. This highlights the distinction between the standard model of measurement and the equilibration paradigm; given temporal control, one can of course generate the desired correlations between the system and the observers, but considering a spontaneous and uncontrolled process, one finds that on average there are no correlations whatsoever between the system and the observers.

\mpel{To further illustrate this distinction, consider a Hamiltonian of the form }
\begin{equation} \label{eHamStandardPlusFree}
\mpel{    H=H_\sys\otimes \mathds{1}_\mathrm{E} + \mathds{1}_\sys\otimes H_\mathrm{E} + X_\sys\otimes Y_\mathrm{E}, }
\end{equation}
\mpel{where $X_\sys$ is diagonal in the eigenbasis of $H_\sys$. Note that Eq.~\eqref{eHamStandardPlusFree} is a necessary but not sufficient condition for the existence of a pointer basis in Quantum Darwinism~\cite{duruisseau2023pointer}. Writing the spectral decompositions $H_\sys=\sum_{i}\varepsilon_{i}\ketbra{i}{i}$ and $X_\sys=\sum_{i}x_{i}\ketbra{i}{i}$, this Hamiltonian can then be rewritten as}
\begin{equation}
\mpel{    H=\sum_{i}\ketbra{i}{i}\otimes H_\mathrm{E}^{(i)}, }
\end{equation}
\mpel{where $H_\mathrm{E}^{(i)}:=\varepsilon_{i}\mathds{1}_\mathrm{E}+ x_{i}Y_\mathrm{E} + H_\mathrm{E}$. As shown in \PRLapp{appsec:constraining}, the above Hamiltonian, in contrast to the standard model of measurement, can in fact lead to correlations between the system and the environment in the pointer basis (in this case, the eigenbasis of the system's free Hamiltonian). Thus, in the equilibration paradigm, the free evolution of either the system, the environment, or both, are in fact necessary in order for the stable encoding of any measurement information into the environment.}

\section{Approximate projective measurements} \label{sec:ApproximateSBS}
While exact equilibration to a state with Spectrum Broadcast Structure is impossible, there remains the question of how well an equilibrium state can approximate this structure. We saw earlier that, with increasing dimension of $\bigotimes_{k=1}^{N}\mathcal{H}_{k}$, one can approach the desired orthogonality condition required for objectivity (though recall that this was a weaker condition than Spectrum Broadcast Structure). In the same vein, we will now see that by collecting the observers' systems together into composite systems, which we call macro-observers, one can approach Spectrum Broadcast Structure exponentially as the size of these macro-observers increases.

As discussed above, equilibration to a state exhibiting the appropriate statistics in the pointer basis, with no coherences between different pointer states, and allowing correlations between the observers and the system in this basis requires a conditional structure of the Hamiltonian given in Eq.~\eqref{eConditionalForm}. With the Strong Independence condition in mind, which demands that the observers should be uncorrelated except via their correlations with the system, we expand this to consider an Hamiltonian of the form
\begin{equation} \label{eq:InteractionHamiltonianAprr}
     H = \sum_{i} \ketbra{i}{i}_\sys \otimes \sum_{k=1}^{N} c_k H_{k}^{(i)}. 
\end{equation}
\mpel{This can be obtained from Eq.~\eqref{eHamStandardPlusFree}, for example, by assuming that both $H_\mathrm{E}$ and $V_\mathrm{E}$ consist of a sum of terms each acting only non-trivially on a single environmental subsystem $k$.} We can see Eq.~\eqref{eq:InteractionHamiltonianAprr} as coupling $k$ to $\sys$ with coupling strength $c_k$, such that for the system state~$i$, the action $H_{k}^{(i)}$ is performed. Assuming $ H$ in Eq.~\eqref{eq:InteractionHamiltonianAprr} to be non-degenerate and starting with the state ${\rho(0)=\rho_{\sys,0}\bigotimes_{k=1}^N \rho_{k,0}}$, we arrive at the equilibrium state
\begin{align} \label{eRhoInfApprox}
    \rho_\mathrm{eq}=\sum_{i=1}^{d_\sys} p_i \, \ketbra{i}{i}_\sys \bigotimes_{k=1}^{N} \rho_k^{(i)},
\end{align}
where $\rho_k^{(i)}$ are obtained from $\rho_{k,0}$ by applying the pinching map with respect to $H_{k}^{(i)}$. Note that the form of this state is invariant under the partial trace with respect to any subset of observers. The state in Eq.~\eqref{eRhoInfApprox} has the form required for Spectrum Broadcast Structure, but it remains to check the orthogonality condition. To that end, let $\ket{\varepsilon_{m,k}^{(i)}}$ denote the eigenvector of $H_{k}^{(i)}$ corresponding to the eigenvalue $\varepsilon_{m,k}^{(i)}$. For simplicity of presentation, we assume that $H_{k}^{(i)}$ and $H_{k}^{(j)}$ have no common eigenvectors (in \PRLapp{appsec:typicality} we show how to obtain a similar result without this assumption). We will find it useful to define the quantity
\begin{equation} \label{eEtaIJ}
    \eta_k^{(ij)}:=\sum_{m}\sqrt{\bra{\varepsilon_{m,k}^{(i)}} \rho_k^{(i)} \rho_k^{(j)} \ket{\varepsilon_{m,k}^{(i)}}} .
\end{equation}

\begin{figure}[t!]
\begin{center}
\includegraphics[width=0.8\columnwidth]{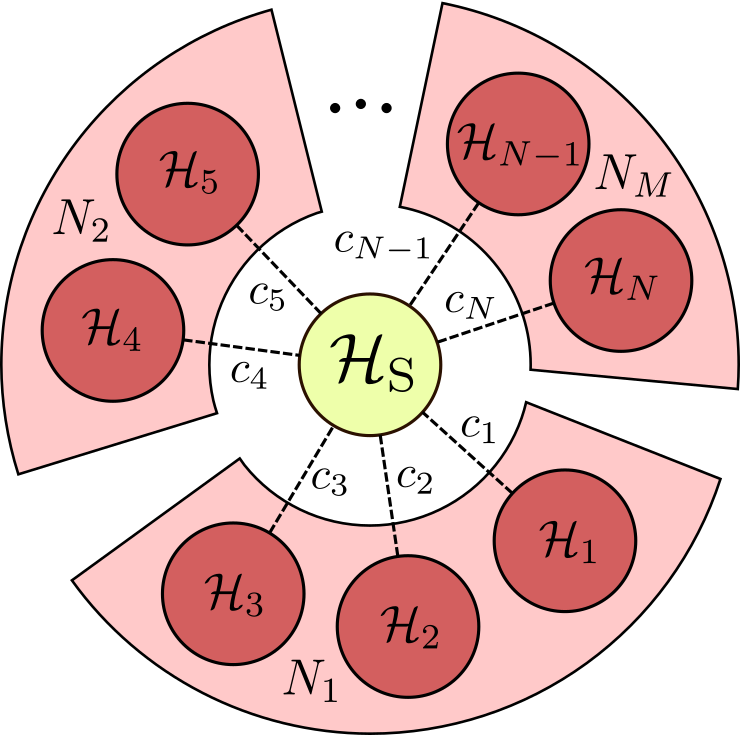}
\caption{A schematic representation of the conditional Hamiltonian given in Eq.~\eqref{eq:InteractionHamiltonianAprr}. It depicts the interaction of system $\mathcal{H}_\mathrm{S}$ (in yellow) with individual parts of the environment $\mathcal{H}_k$ (in red). The coupling strength between the system and each environment is indicated by the dashed lines and the labels $c_k$. The light-red frames exemplify ways of coarse-graining the environments into macro-observers $N_k$.}
\label{fig:HamiltoianStructure}
\end{center}
\end{figure}

Now, let $M<N$ denote the number of macro-observers, and let $N_{q}$ denote the set of $k$ labels to be collected together to form a given macro-observer $q$ (see Fig.~\ref{fig:HamiltoianStructure}).  We can then write Eq.~\eqref{eRhoInfApprox} in terms of macro-observers as $\rho_\mathrm{eq}=\sum_{i=1}^{d_\sys} p_i \, \ketbra{i}{i}_\sys \bigotimes_{q=1}^{M} \tilde{\rho}_q^{(i)}$, where $\lbrace\tilde{\rho}_q^{(i)}:=\bigotimes_{k\in N_q}\rho_k^{(i)}\rbrace$ are states on $\bigotimes_{k\in N_q} \mathcal{H}_{k}$. As we show in \PRLapp{appsec:constraining}, the fidelity between the conditional states of the macro-observers satisfies
\begin{equation}
    F(\tilde{\rho}_q^{(i)},\tilde{\rho}_q^{(j)}) \leq e^{- \gamma_{q}^{(ij)} \vert N_{q}\vert} ,
\end{equation}
where 
\begin{equation}
    \gamma_{q}^{(ij)} := -2 \ln \left(\max_{k \in N_{q}} \eta_k^{(ij)}\right) > 0,
\end{equation}
Thus as we increase each $\vert N_{q}\vert$, \textit{i.e.} the the number of degrees of freedom collected together to form each macro-observer $q$, the equilibrium state tends exponentially towards one with Spectrum Broadcast Structure for the macro-observers. In other words, as we coarse-grain our view of observers towards ever-larger collections of microscopic systems, we tend exponentially towards a state of objectivity commensurate with the occurrence of a measurement; only sufficiently macroscopic observers will see objective facts about the system.

\section{Discussion and conclusions} \label{sec:Discussion}

Quantum measurement is often described as the `collapse of the wavefunction' when a quantum system makes contact with an observer. One is then led to the question, what exactly can constitute an observer?~\cite{brukner_2015}. The Measurement-Equilibration Hypothesis allows us to constrain this question, giving the answer that an observer is whatever system interacts with the system of interest such that the result equilibrates to a state that approximates objective state structure asymptotically. In this paradigm, we have shown the impossibility of exact projective measurements (in accordance with known results~\cite{Guryanova_2020}). However, they can be approximated exponentially as more and more microscopic systems are incorporated into a given macroscopic observer. This accords with the intuition that small systems should not be considered observers~\cite{brukner2021qubits}, but that larger ones can be~\cite{Brukner2022,Wiseman2022a}. 

\mpel{Moreover, this hypothesis provides a mechanism by which decoherence can arise not only through the usual mechanism of tracing out environmental degrees of freedom, but also intrinsically within isolated systems, due to the limited resolving power of equilibrating observables. Any observable not capable of resolving the continued microscopic evolution of the quantum state becomes likewise insensitive to the residual quantum coherence between distinct elements of the measurement basis; objective behavior arises in a closed, unitary setting, in a spontaneous and irreversible manner, robust to all but the most fine-tuned initial conditions.}

We stress that we have not sought to address the problem of which state an observer sees (the so-called ``small'' measurement problem~\cite{brukner_2015}), nor to explain how the Born rule comes about. Instead, we have adopted a physical principle (in the form of a hypothesis) and asked if it can explain how observers may, through unitary dynamics, spontaneously and irreversibly become correlated with the system under investigation, and if this can occur such that multiple observers will agree on the measurement outcome. We have not explicitly adopted a given interpretation of quantum theory, and the compatibility (or lack thereof) of different interpretations with the Measurement-Equilibration Hypothesis is an interesting question which we leave for future work.

The results presented here can be applied to the spin-spin model studied in~\cite{Mironowicz2017}, providing a physical interpretation of its long-time objectivity. Furthermore, understanding the initially uncorrelated state as an out-of-equilibrium resource, the Measurement-Equilibration Hypothesis provides a novel perspective on engines based on the extraction of work via quantum measurements~\cite{elouard2017extracting,elouard2018efficient}, and thus a potential path to resolving the foundational questions regarding the source of this energy~\cite{jordan2020quantum}.

In the simple version of an equilibrating measurement presented in this article, the Strong Independence condition on the equilibrium state led to the demand that observers interact via a Hamiltonian with a conditional form, and surprisingly, that the particular conditional form given by the standard model of measurement does not satisfy the most minimal requirement for measurement, namely correlation of observer and system. Thus, while an average over such Hamiltonians appears to result in objectivity at given times~\cite{korbicz2017generic}, any given Hamiltonian of the ensemble does not equilibrate on average to an objective one. \mpel{Including free evolution of either the system or the environment, on the other hand, ensures the stable encoding of measurement information into the environment.} The conditional form imposed by Strong Independence does not resemble the usual forms of interaction found in nature, as exemplified by example in Eq.~\eqref{eConditionalForm}, suggesting that the model needs to be expanded upon if the Measurement-Equilibration Hypothesis is to be employed in the modelling of real-world experiments. Nonetheless, our results suggest a potential resolution to the question of when a unitary description of dynamics might give way to a projective description~\cite{brukner_2015}: when the system of interest interacts conditionally and independently with a number of distinct systems, in such a way that the whole system equilibrates. 

In this regard, two immediate generalisations may suffice. First, one might allow for vanishingly small correlations violating the Strong Independence condition. It seems likely that the requirement that such violations be vanishingly small will nonetheless strongly constrain the form of the interaction. A second, and in our view more promising generalisation, would be the inclusion of an initial ``pre-measurement'' step, whereby one allows some time-controlled interaction to correlate an apparatus with the system of interest, whose subsequent equilibration with some observers leads to a state of objectivity between the system of interest and the observers; in this case, the requirement that the system of interest be undisturbed in the measurement basis would be less restrictive in the equilibrium step. Indeed many measurements modelled via projection operators do not leave the system in an eigenstate corresponding to the measurement outcome, \textit{e.g.}~photon detection. \mpel{Alternatively, one may relax the requirement of objective features at full equilibrium, and instead seek an intermediate ``measured'' phase as a prethermalisation phenomenon~\cite{berges2004prethermalization}, whereby a system quickly reaches a long-lived pseudo-equilibrium, before finally relaxing to true equilibrium. Such a picture would generalise the observation in~\cite{riedel2012rise}.}

The paradigm introduced here provides a new thermodynamically consistent framework for tackling questions regarding the role of the preparation/driving of measurement apparatuses and the characterisation of the necessary out-of-equilibrium resources in the measurement process. Since the equilibrium states considered are not thermal, it is of interest to examine the relationship between our model and many-body localisation~\cite{abanin2017} and in particular whether a pre-thermalisation phenomenon~\cite{Gring2012} may be at play. This is reinforced by \cite{le2021thermality}, which shows that thermality and objectivity are often at odds, showing the need for a more general equilibration approach. Indeed, the form of Hamiltonian required to equilibrate to a classical-quantum state, Eq.~\eqref{eConditionalForm}, imposes the conservation of any observable that is diagonal in the pointer basis, thus preventing local thermalisation with respect to that observable. 

In addition, one may go beyond the simple statement of the measurement-equilibration hypothesis above to investigate whether the quality of the measurement depends on the amount of entropy dissipated in the process, and how this may relate to the timescale of the measurement. This is of interest when one considers the role of observables, which we have not done here. Macroscopic observables are in general highly degenerate, and therefore tend to equilibrate (see \textit{e.g.}~\cite{Anza2018}), and it may be fruitful to consider if this pertains to the very high speed associated with a projective measurement~\cite{strasberg2020how}, particularly in light of recent work progress relating an increase in entropy to the degeneracy of observables~\cite{meier2024emergence}. If the resulting equilibration timescales can be faster than the free dynamics of a system of interest, this may resolve the question raised in Sec.~\ref{sExactImpossible} regarding measurements which do not commute with generator of those free dynamics.

While ours is only a first step towards a thermodynamically self-contained model for quantum measurements, we hope that this will eventually resolve many of the foundational issues that manifest through the apparent inconsistency of textbook postulates of quantum measurements and thermodynamic laws. 

\section{Acknowledgements} The authors thank Nicolai Friis, Erickson Tjoa and Faraj Bakhshinezhad for helpful discussions, and in particular Sophie Engineer for suggesting that free Hamiltonians may permit correlations in the standard model of measurement. They further thank Alessandro Candeloro and Tom Rivlin for insightful comments on the manuscript, and Nick Ormrod for highlighting a flaw in an earlier version of an appendix. This project/ publication was made possible through the support of Grant 62423 from the John Templeton Foundation. The opinions expressed in this publication are those of the author(s) and do not necessarily reflect the views of the John Templeton Foundation. E.S. acknowledges support from the Austrian Science Fund (FWF) through the START project Y879-N27, from the ESQ Discovery grant ``Emergence of physical laws: From mathematical foundations to applications in many body physics'', and from the European flagship on quantum technologies (‘ASPECTS’ consortium 101080167), F.C.B. acknowledges support by grant number FQXi-RFP-IPW-1910 from the Foundational Questions Institute and Fetzer Franklin Fund, a donor advised fund of Silicon Valley Community Foundation; and Irish Research Council Laureate Award IRCLA/2022/3922. E.S., M.H. and M.P.E.L. acknowledge funding from the European Research Council (Consolidator grant ``Cocoquest'' 101043705). 


\onecolumngrid
\appendix

\section{The impossibility of ideal projective measurements}\label{appsec:constraining}
In this section, we demonstrate the impossibility of performing an exact projective measurement according to the Measurement Equilibration Hypothesis. In particular, we prove that an initially uncorrelated state cannot equilibrate to one with Spectrum Broadcast Structure; this follows from a stronger result, described in the following. Firstly, we allow the possibility of initial correlations between the observers' systems, but not with the system to be measured i.e. $\rho(0)=\rho_{\sys,0}\otimes\rho_{\mathrm{E},0}$, where $\rho_{\sys,0}$ is the initial state of the system and $\rho_{\mathrm{E},0}$ the initial state of the environment. Secondly, we demand equilibration to a classical-quantum state~\cite{devetak2004distilling} with respect to the measurement basis, with the appropriate probabilities, \textit{i.e.} a state of the form
\begin{align} \label{eClassQuantState}
    \rho_{\text{CQ}} = \sum_i p_i\ketbra{i}{i}_\sys\otimes \rho_\mathrm{E}^{(i)}
\end{align}
for some $\rho_\mathrm{E}^{(i)}$ on $\mathcal{H}_\mathrm{E}$. Thirdly, we demand orthogonality of the conditional states $\lbrace\rho_\mathrm{E}^{(i)}\rbrace_i$, \textit{i.e.} $\rho_\mathrm{E}^{(i)}\rho_\mathrm{E}^{(j)}=0$, or equivalently $F(\rho_\mathrm{E}^{(i)},\rho_\mathrm{E}^{(j)})=0$, for $i\neq j$. These are weaker requirements than for Spectrum Broadcast Structure, as can be seen by comparing Eqs.~\eqref{eClassQuantState} and~\eqref{eSBSdefn}. Thus when we prove that they cannot be simultaneously satisfied, the impossibility of equilibration to a state with Spectrum Broadcast Structure follows as a consequence.

The proof proceeds by first constraining the form of the Hamiltonian that would be necessary to equilibrate to a classical-quantum state, and then showing that the resulting fidelity between the conditional states $\lbrace\rho_\mathrm{E}^{(i)}\rbrace_i$ is necessarily non-zero.

\subsection{Constraining the Hamiltonian} \label{sConstrainH}

Recalling that if a system equilibrates on average, it equilibrates to its infinite-time average~\cite{gogolin2016equilibration}, our first step is to find which Hamiltonians can give rise to an infinite-time average of the required clasical-quantum form, \textit{i.e.} Eq.~\eqref{eClassQuantState}. Since the infinite-time average is also given by the pinching map with respect to the Hamiltonian, i.e. Eq.~\eqref{ePinch}, it will commute with the Hamiltonian:  
$[\rho_\mathrm{eq} ,  H ]=0$. Consequently, if $\rho_\mathrm{eq}=\rho_\mathrm{CQ}$, then $[\rho_\mathrm{CQ} ,  H ]=0$. We now seek the constraints that the latter equation imposes on the form of $ H$. 

First, noting that the $[\rho_\mathrm{CQ} ,  H ]=0$ should hold for all $\lbrace p_{i}\rbrace$ in Eq.~\eqref{eClassQuantState}, we see that in fact the following condition must hold
\begin{equation} \label{eCommH}
    \left[ \ketbra{i}{i} \otimes\rho_\mathrm{E}^{(i)}, H\right] = 0 \quad \forall \, i .
\end{equation}
\mpel{Let us now impose orthogonality of the conditional states, i.e. $\rho_\mathrm{E}^{(i)}\rho_\mathrm{E}^{(j)}=0$ for $i\neq j$. This condition implies the existence of a basis of $\mathcal{H}_\mathrm{E}$ which simultaneously diagonalises all of the $\lbrace\rho_\mathrm{E}^{(i)}\rbrace_i$. Denoting such a basis by $\lbrace\ket{\lambda}_\mathrm{E}\rbrace_\lambda$, the orthogonality condition implies that we can partition the set of labels $\lambda$ according to which of the conditional states $\lbrace\rho_\mathrm{E}^{(i)}\rbrace_{i}$ has support on $\ket{\lambda}_\mathrm{E}$. In other words, for each $i$ we can define $S_{i}:=\lbrace\lambda \,\vert\,\rho_\mathrm{E}^{(i)}\ket{\lambda}_\mathrm{E}\neq 0 \rbrace$, and then write}
\begin{equation}
    \rho_\mathrm{E}^{(i)}=\sum_{\lambda \in S_{i}} \rho_\lambda^{(i)}\ketbra{\lambda}{\lambda}
\end{equation}
\mpel{for some $\lbrace\rho_\lambda^{(i)}\rbrace_{i,\lambda}$. We can then define the set of projectors}
\begin{equation}
    \Pi^{(i)}_{\sys ,\mathrm{E}} := \sum_{\lambda\in S_{i}} \ketbra{i}{i}_\sys\otimes\ketbra{\lambda}{\lambda}_\mathrm{E}
\end{equation}
\mpel{and then decompose the identity on $\mathcal{H}_\sys\otimes\mathcal{H}_\mathrm{E}$ in terms of the sum of these projectors, plus a projector onto the rest of the Hilbert space}
\begin{equation}
    \mathds{1}_\sys \otimes\mathds{1}_\mathrm{E} = \sum_{i} \Pi^{(i)}_{\sys ,\mathrm{E}} + \Pi_\mathrm{rest} .
\end{equation}
\mpel{Multiplying $H$ from both sides by this resolution of the identity,}
\begin{equation}
\begin{aligned}  \label{eHamDecomp}  
    H &= \sum_{ij} \Pi^{(i)}_{\sys ,\mathrm{E}}H \Pi^{(j)}_{\sys ,\mathrm{E}} + \Pi_\mathrm{rest}H\sum_{i}\Pi^{(i)}_{\sys ,\mathrm{E}} + \sum_{i}\Pi^{(i)}_{\sys ,\mathrm{E}} H \Pi_\mathrm{rest} + \Pi_\mathrm{rest}H\Pi_\mathrm{rest}  \\
    &= \sum_{i} \Pi^{(i)}_{\sys ,\mathrm{E}}H \Pi^{(i)}_{\sys ,\mathrm{E}} + \sum_{i\neq j} \Pi^{(i)}_{\sys ,\mathrm{E}}H \Pi^{(j)}_{\sys ,\mathrm{E}} + \sum_{i}\Pi_\mathrm{rest}H\Pi^{(i)}_{\sys ,\mathrm{E}} + \sum_{i}\Pi^{(i)}_{\sys ,\mathrm{E}} H \Pi_\mathrm{rest} + \Pi_\mathrm{rest}H\Pi_\mathrm{rest}   \\
    &\equiv \sum_{i} H^{(ii)} + \sum_{\substack{i,j \\ i\neq j}} H^{(ij)} + \sum_{i} H^{(\mathrm{rest},i)} + \sum_{i} H^{(i,\mathrm{rest})} + H_\mathrm{rest},
\end{aligned}
\end{equation}
\mpel{we thus separate it into a sum of linearly independent terms: those acting only on the subspace corresponding to an outcome $i$, i.e.~$H^{(ii)}:=\Pi^{(i)}_{\sys ,\mathrm{E}}H \Pi^{(i)}_{\sys ,\mathrm{E}}$, those mapping from the subspace corresponding to an outcome $j$ to the one corresponding to an outcome $i\neq j$, i.e.~$H^{(ij)}:=\Pi^{(i)}_{\sys ,\mathrm{E}}H \Pi^{(j)}_{\sys ,\mathrm{E}}$, those mapping from the space corresponding to an outcome $i$ to the ``rest'' part of the space, i.e.~$H^{(\mathrm{rest},i)}$, those mapping in the other direction, i.e.~$H^{(i,\mathrm{rest})}$, and one acting only on the ``rest'' part of the space, i.e.~$H_\mathrm{rest}$. With this decomposition, the condition in Eq.~\eqref{eCommH} now becomes}
\begin{equation} 
    \ketbra{i}{i} \otimes\rho_\mathrm{E}^{(i)} \left( H^{(ii)} + \sum_{j\neq i} H^{(ij)}  + H^{(i,\mathrm{rest})} \right) - \left(  H^{(ii)} + \sum_{j\neq i} H^{(ji)}  + H^{(\mathrm{rest},i)}  \right) \ketbra{i}{i} \otimes\rho_\mathrm{E}^{(i)} = 0 \quad \forall \, i ,
\end{equation}
\mpel{i.e.}
\begin{equation} 
    [\ketbra{i}{i} \otimes\rho_\mathrm{E}^{(i)} , H^{(ii)} ]+
    \ketbra{i}{i} \otimes\rho_\mathrm{E}^{(i)} \left(  \sum_{j\neq i} H^{(ij)}  + H^{(i,\mathrm{rest})} \right) - \left( \sum_{j\neq i} H^{(ji)}  + H^{(\mathrm{rest},i)}  \right) \ketbra{i}{i} \otimes\rho_\mathrm{E}^{(i)} = 0 \quad \forall \, i .
\end{equation}
\mpel{Noting that each term is linearly independent (counting the commutator as a single term), we see that they must individually vanish, i.e.}
\begin{align} 
    [ \ketbra{i}{i} \otimes\rho_\mathrm{E}^{(i)} , H^{(ii)} ] &= 0 \quad \forall \, i, \\
    \ketbra{i}{i} \otimes\rho_\mathrm{E}^{(i)} H^{(ij)} &= 0 \quad \forall \, i, \label{eCommH1} \\
    \ketbra{i}{i} \otimes\rho_\mathrm{E}^{(i)} H^{(i,\mathrm{rest})} &= 0 \quad \forall \, i, \label{eCommH2} \\
    H^{(ji)} \ketbra{i}{i} \otimes\rho_\mathrm{E}^{(i)} &= 0 \quad \forall \, i, \label{eCommH3} \\
    H^{(\mathrm{rest},i)} \ketbra{i}{i} \otimes\rho_\mathrm{E}^{(i)} &= 0 \quad \forall \, i . \label{eCommH4}
\end{align}
\mpel{Recall that $H^{(ij)}$ and $H^{(i,\mathrm{rest})}$ are by definition the parts of $H$ that map to the support of $\ketbra{i}{i} \otimes\rho_\mathrm{E}^{(i)}$, and similarly $H^{(ji)}$ and $H^{(\mathrm{rest},i)}$ are by definition the parts of $H$ that map from the support of $\ketbra{i}{i} \otimes\rho_\mathrm{E}^{(i)}$. Thus the only way for Eqs.~\eqref{eCommH1}-\eqref{eCommH4} to be satisfied is if $H^{(ij)}=0', \forall \, i,j,i\neq j$, and $H^{(\mathrm{rest},i)}=H^{(i,\mathrm{rest})}=0 \,\forall \, i$, and therefore the condition in Eq.~\eqref{eCommH} implies that}
\begin{equation} 
    H =  \sum_{i} H^{(ii)} + H_\mathrm{rest},
\end{equation}
\mpel{and inserting the definition of $H^{(ii)}$, we can write this as}
\begin{equation} \label{eHamFormWRest}
    H =\sum_{i} \ketbra{i}{i}_\sys\otimes H_\mathrm{E}^{(i)} + H_\mathrm{rest},
\end{equation}
\mpel{where}
\begin{equation}
    H_\mathrm{E}^{(i)} := (\bra{i}_\sys \otimes\mathds{1}_\mathrm{E}) H (\ket{i}_\sys \otimes\mathds{1}_\mathrm{E})
\end{equation}
\mpel{has support only on $\lbrace \ket{\lambda} \rbrace_{\lambda\in S_{i}}$.}

\mpel{Recalling that $\rho_\mathrm{CQ}$, and therefore by assumption $\rho_\mathrm{eq}$, has zero support on the ``rest'' subspace of $\mathcal{H}_\sys\otimes\mathcal{H}_\mathrm{E}$, and noting that the Hamiltonian in Eq.~\eqref{eHamFormWRest} cannot map between the ``rest'' subspace and its complement, we see that the initial state must likewise have no support on this subspace, i.e.}
\begin{equation}
    H_\mathrm{rest} \rho_{\sys,0}\otimes\rho_{\mathrm{E},0} = \rho_{\sys,0}\otimes\rho_{\mathrm{E},0}  H_\mathrm{rest} = 0 .
\end{equation}
\mpel{In other words, for the equilibrium state to have the classical-quantum form $\rho_\mathrm{CQ}$, it must either be the case that $H_\mathrm{rest}=0$, or that the support of the initial state is fully contained within the kernel of $H_\mathrm{rest}$.}

\mpel{Since the $H_\mathrm{rest}$ term in the Hamiltonian in Eq.~\eqref{eHamFormWRest} commutes with the other terms, the time evolution under $H$ factorizes as}
\begin{align}
    e^{-i  Ht} = e^{-i \sum_{i} \ketbra{i}{i}_\sys\otimes H_\mathrm{E}^{(i)} t} e^{-i  H_\mathrm{rest} t}.
\end{align}
\mpel{and the initial state then evolves as}
\begin{equation}
\begin{aligned}
   e^{-i  Ht}\rho_{\sys,0}\otimes\rho_{\mathrm{E},0}e^{i  Ht} &= e^{-i \sum_{i} \ketbra{i}{i}_\sys\otimes H_\mathrm{E}^{(i)} t} e^{-i  H_\mathrm{rest} t} \rho_{\sys,0}\otimes\rho_{\mathrm{E},0} e^{i  H_\mathrm{rest} t}e^{i \sum_{i} \ketbra{i}{i}_\sys\otimes H_\mathrm{E}^{(i)} t} \\
   &= e^{-i \sum_{i} \ketbra{i}{i}_\sys\otimes H_\mathrm{E}^{(i)} t} \rho_{\sys,0}\otimes\rho_{\mathrm{E},0} e^{i \sum_{i} \ketbra{i}{i}_\sys\otimes H_\mathrm{E}^{(i)} t} .
\end{aligned}
\end{equation}

\mpel{Since the ``rest'' part of the Hamiltonian, if non-zero, cannot have any effect on the dynamics, we can set it to zero in the following without loss of generality, taking}
\begin{equation} \label{eHamForm}
    H = \sum_{i} \ketbra{i}{i}_\sys\otimes H_\mathrm{E}^{(i)} .
\end{equation} 
This generates unitary evolution of the form 
\begin{align}
    e^{-i  Ht} = \sum_{i} \ketbra{i}{i} \otimes U_\mathrm{E}^{(i)}(t),
\end{align}
where $U_\mathrm{E}^{(i)}(t) := \exp \left[ -i H_\mathrm{E}^{(i)}t  \right]$, which this follows directly from the power series definition of the operator exponential and orthonormality of the states $\lbrace\ket{i}\rbrace$.
We can then write the equilibrium state in the form of the infinite time average ${\langle \rho(t) \rangle_\infty:= \lim_{T\rightarrow \infty}\tfrac{1}{T}\int_0^T dt e^{-i H t} \rho(0)e^{i Ht}}$ as
\begin{align}\label{eq:rhoinftensorstructure}
    &\rho_\mathrm{eq} = \Big\langle \sum_i \ketbra{i}{i}\otimes U_\mathrm{E}^{(i)}(t) \rho_{\mathrm{S},0}\otimes\rho_{\mathrm{E},0} \sum_j \ketbra{j}{j}\otimes U_\mathrm{E}^{(j)\dagger}(t)\Big\rangle_\infty\nonumber\\&
    =\sum_{i,j} \bra{i}\rho_{\mathrm{S},0}\ket{j} \ketbra{i}{j}\otimes \Big\langle U_\mathrm{E}^{(i)}(t)\rho_{\mathrm{E},0}U_\mathrm{E}^{(j)\dagger}(t)\Big\rangle_\infty \stackrel{!}{=}\sum_i p_i \ketbra{i}{i}\otimes \rho_\mathrm{E}^{(i)}
\end{align}
which implies that
\begin{align} \label{eEnvAverage}
    \Big\langle U_\mathrm{E}^{(i)}(t)\rho_{\mathrm{E},0}U_\mathrm{E}^{(j)\dagger}(t)\Big\rangle_\infty &\stackrel{!}{=} \, \begin{cases} 0 & i\neq j \\
  \rho_\mathrm{E}^{(i)} & i=j . \end{cases}
\end{align}  
Now, let us denote the eigenvalues of $ H_\mathrm{E}^{(i)}$ by $\lbrace \varepsilon_{n}^{(i)} \rbrace_{n}$, with associated eigenvectors $\lbrace \ket{\varepsilon_{n}^{(i)}} \rbrace_{n}$. Writing $U_\mathrm{E}^{(i)}(t)$ and $U_\mathrm{E}^{(j)}(t)$ in the left hand side of Eq.~\eqref{eEnvAverage} in the eigenbasis of $ H_\mathrm{E}^{(i)}$ and $ H_\mathrm{E}^{(j)}$ respectively, we have
\begin{equation}
     \Big\langle U_\mathrm{E}^{(i)}(t)\rho_{\mathrm{E},0}U_\mathrm{E}^{(j)\dagger}(t)\Big\rangle_\infty = \sum_{n,m} \Big\langle e^{-i(\varepsilon_{n}^{(i)}-\varepsilon_m^{(j)})t}\Big\rangle_\infty \bra{\varepsilon_{n}^{(i)}} \rho_{\mathrm{E},0} \ket{\varepsilon_m^{(j)}} \ketbra{\varepsilon_{n}^{(i)}}{\varepsilon_m^{(j)}} ,
\end{equation}
and so in order to satisfy Eq.~\eqref{eEnvAverage}, we must have that 
\begin{equation}
    \Big\langle e^{-i(\varepsilon_{n}^{(i)}-\varepsilon_m^{(j)})t}\Big\rangle_\infty \bra{\varepsilon_{n}^{(i)}} \rho_{\mathrm{E},0} \ket{\varepsilon_m^{(j)}} \stackrel{!}{=} 0 \quad \forall  i,j,n , m\; \mathrm{where}\; i\neq j .
\end{equation}
Let write the spectral decomposition of the conditional Hamiltonians as $H_\mathrm{E}^{(i)}=\sum_{n} \varepsilon_{n}^{(i)}\Pi_n^{(i)}$, where $\lbrace\Pi_n^{(i)}\rbrace_{n}$ are projectors onto degenerate sub-spaces of $H_E^{(i)}$ (or onto its eigenvectors if it is not degenerate). If the $\lbrace H_\mathrm{E}^{(i)}\rbrace_{i}$ have no eigenvalues in common, \textit{i.e.}~$\varepsilon_{n}^{(i)}\neq\varepsilon_m^{(j)}$ $\forall i,j,n,m$ with $i\neq j$, then the above equation holds independently of the initial state $\rho_{\mathrm{E},0}$ and we arrive at a state of the form
\begin{align} \label{ecqs}
        \rho_\mathrm{eq} = \sum_i p_i \ketbra{i}{i}\otimes \rho_\mathrm{E}^{(i)} \quad \text{with} \quad \rho_\mathrm{E}^{(i)}= \sum_{n} \Pi_n^{(i)} \rho_{\mathrm{E},0} \Pi_n^{(i)}.
\end{align}
If this is not the case, \textit{i.e.} when there are values of $\varepsilon_{n}^{(i)}$ and $\varepsilon_m^{(j)}$ that coincide (\emph{i.e.} degeneracies of $H$), then we can split $\rho_\mathrm{E}^{(i)}$ into two terms in the following way:
\begin{align} \label{edegeneracycqt}
       \rho_\mathrm{eq} = \sum_{i} p_i \ketbra{i}{i}\otimes \sum_{n} \Pi_n^{(i)} \rho_{\mathrm{E},0} \Pi_n^{(i)}  + \sum_{\substack{ijnm \\ \varepsilon_{n}^{(i)}=\,\varepsilon_{m}^{(j)}, \, i\neq j}}  \bra{i}\rho_{\mathrm{S},0}\ket{j} \ketbra{i}{j}\otimes\bra{\varepsilon_{n}^{(i)}} \rho_{\mathrm{E},0} \ket{\varepsilon_{m}^{(j)}} \ketbra{\varepsilon_{n}^{(i)}}{\varepsilon_{m}^{(j)}} .
\end{align}
Then, for Eq.~\eqref{eEnvAverage} to hold we must have $\bra{\varepsilon_{n}^{(i)}} \rho_{\mathrm{E},0} \ket{\varepsilon_m^{(j)}}=0$ for each pair of coinciding eigenvalues, $\varepsilon_{n}^{(i)}=\varepsilon_{m}^{(j)}$ $i\neq j$, requiring an initially rank-deficient state, in which case the second sum in Eq.~\eqref{edegeneracycqt} vanishes and we are left with a state of the form given in Eq.~\eqref{ecqs}, as required.
As an aside, we note that this state satisfies the faithfulness requirement in~\cite{Guryanova_2020} (i.e.: $\sum_i\tr[\ketbra{i}{i}\otimes\Pi_i\rho_\mathrm{eq}]=1$ for orthogonal projectors $\Pi_i$ with $[\Pi_i,\rho_E^{(i)}]=0$). 

We have so far made no demands on the microscopic structure on $\mathcal{H}_\mathrm{E}$, aside from the potential rank-deficient initial states required to deal with degeneracies in $H$. A Hamiltonian of the form of Eq.~\eqref{eHamForm} is, therefore, necessary for the formation of classical-quantum states in general.

\subsection{Bounding the fidelity between conditional states}
We now show that the orthogonality requirement, \textit{i.e.} $F(\rho_\mathrm{E}^{(i)},\rho_\mathrm{E}^{(j)})=0$ for $i\neq j$, cannot be satisfied exactly. In particular, we show that
\begin{align}\label{eq:fidelitybound}
    F(\rho_\mathrm{E}^{(i)},\rho_\mathrm{E}^{(j)})\geq \dfrac{\tr(\rho_{\mathrm{E},0}^2)}{d_\mathrm{E}^2} ,
\end{align}
where $d_\mathrm{E}:=\dim(\mathcal{H}_\mathrm{E})$.

First note that the conditional time-averaged state $\rho_\mathrm{E}^{(i)}$ in Eq.~\eqref{ecqs} is a result of the pinching map with respect to the projectors $\lbrace\Pi_n^{(i)}\rbrace_{n}$. Denoting the number of these projectors (\textit{i.e.} the number of distinct eigenvalues of $ H_\mathrm{E}^{(i)}$) by $d^{(i)}$, we can rewrite the pinching map as a mixed-unitary channel \cite{tomamichel2016,watrous2018} as follows
\begin{align}
    \rho_\mathrm{E}^{(i)}=\sum_{n=1}^{d^{(i)}} \Pi_n^{(i)} \rho_{\mathrm{E},0} \Pi_n^{(i)} = \dfrac{1}{d^{(i)}}\sum_{y=1}^{d^{(i)}} U_y^{(i)} \rho_{\mathrm{E},0} U_y^{(i)\dagger} 
\end{align}
where $U_{y}^{(i)} :=\sum_{n=1}^{d^{(i)}} e^{-i\tfrac{ 2\pi n y}{d^{(i)}}} \Pi_n^{(i)}$.
Noting that $U^{(i)}_{d^{(i)}} =\mathds{1}$, we write
\begin{align}
    \rho_\mathrm{E}^{(i)}=\dfrac{\rho_{\mathrm{E},0}}{d^{(i)}}+\dfrac{1}{d^{(i)}}\sum_{y=1}^{d^{(i)}-1} U_y^{(i)} \rho_{\mathrm{E},0} U_y^{(i)\dagger} 
\end{align}
and thus
\begin{align}
    \tr[(\rho_\mathrm{E}^{(i)}\rho_\mathrm{E}^{(j)}]=&\dfrac{1}{d^{(i)}d^{(j)}}\Bigg\lbrace\tr\Big[\rho_{\mathrm{E},0}^2\Big] 
    + \tr\Big[\rho_{\mathrm{E},0}\sum_{y=1}^{d^{(i)}-1} U_y^{(j)} \rho_{\mathrm{E},0} U_y^{(j) \dagger}\Big]
    +\tr\Big[\sum_{y=1}^{d^{(i)}-1} U_y^{(i)} \rho_{\mathrm{E},0} U_y^{(i) \dagger}\rho_{\mathrm{E},0}\Big] \nonumber \\
    &+ \tr\Big[\sum_{y=1}^{d^{(i)}-1} U_y^{(i)} \rho_{\mathrm{E},0} U_y^{(i) \dagger}\sum_{y'=1}^{d^{(i)}-1} U_{y'}^{(j)} \rho_{\mathrm{E},0} U_{y'}^{(j) \dagger} \Big]\Bigg\rbrace .
\end{align}    
Since the trace of the product of two positive operators is itself positive, we can remove the second, third and fourth terms to obtain the inequality
\begin{align}
    \tr(\rho_\mathrm{E}^{(i)}\rho_\mathrm{E}^{(j)}) \geq \dfrac{\tr(\rho_{\mathrm{E},0}^2)}{d^{(i)}d^{(j)}} .
\end{align}
Now, note that the fidelity between two states satisfies $F(\rho,\sigma)\geq \tr(\rho \sigma)$ \cite{jozsa1994}, allowing us to bound the Fidelity
\begin{align} \label{eFidBoundij}
    F(\rho_\mathrm{E}^{(i)},\rho_\mathrm{E}^{(j)})\geq \dfrac{\tr(\rho_{\mathrm{E},0}^2)}{d^{(i)}d^{(j)}} \geq \dfrac{\tr(\rho_{\mathrm{E},0}^2)}{d_\mathrm{E}^2},
\end{align}
since $d_\mathrm{E}\geq d^{(i)}$ $\forall i$. Thus, the fidelity between any two time-averaged states with the same initial state cannot be zero. It is therefore impossible to equilibrate exactly to a classical-quantum state, Eq.~\eqref{eClassQuantState}, satisfying the orthogonality condition $F(\rho_\mathrm{E}^{(i)},\rho_\mathrm{E}^{(j)})=0$. As a consequence, it is impossible for an initially uncorrelated state to equilibrate exactly to a state with Spectrum Broadcast Structure, as this is a stronger condition, as discussed above.

\section{The standard measurement model does not equilibrate to a correlated state}\label{appsec: SBS-unattainable with common Hamiltonian}
In this section we consider the Hamiltonian used in the von-Neumann measurement scheme (or the ``standard model of measurement'')~\cite{Busch_1996}:
\begin{align} \label{eVonNeuHam}
    H=X_\mathrm{S}\otimes Y_{\mathrm{E}} ,
\end{align}
where $X_\mathrm{S}$ is diagonal in the measurement basis, $X_\mathrm{S}=\sum_i x_i \ketbra{i}{i}_\mathrm{S}$. By carefully controlling the time for which this Hamiltonian is applied, one can correlate the system S with the environment E in this basis. However, without temporal control -- specifically in an equilibration paradigm -- this Hamiltonian leads to an equilibrium state that does not feature any correlations between system and environment, as we now show.

The unitary operator generated by Eq.~\eqref{eVonNeuHam} is given by
\begin{align}
    e^{-\mathrm{i} H t}= e^{-i X_\mathrm{S}\otimes Y_\mathrm{E} t}=e^{-i(\sum_i x_i \ketbra{i}{i}_\mathrm{S}\otimes Y_\mathrm{E})t}= \sum_i\ketbra{i}{i}_\mathrm{S}\otimes e^{-i x_i Y_\mathrm{E} t},
\end{align}
As in the previous section, we assume that system and environment are initially uncorrelated, such that the initials state is given by
\begin{align}
    \rho(0)=\rho_{\mathrm{S},0}\otimes\rho_{_\mathrm{E},0},
\end{align}
and we calculate the time-evolved state of the total system
\begin{align} \label{eVonNeuEvolvedState}
    \rho(t)= e^{-i Ht}\rho(0)e^{i Ht}= e^{-i X_\mathrm{S}\otimes Y_\mathrm{E} t}\Big(\rho_\mathrm{S,0}\otimes \rho_{\mathrm{E},0}\Big) e^{i X_\mathrm{S}\otimes Y_\mathrm{E} t} =\sum_{i,j}p_{ij}\ketbra{i}{j}\otimes \rho_\mathrm{E}^{(ij)}(t)
\end{align}
where $p_{ij}$ are the elements of the initial system state in the pointer basis, \emph{i.e.} $\rho_{\mathrm{S},0}=\sum_{ij}^{d_s} p_{ij} \ketbra{i}{j}$, and we have defined $\rho_\mathrm{E}^{(ij)}(t):=e^{-ix_i Y_\mathrm{E} t}\rho_{\mathrm{E},0} e^{ix_i Y_\mathrm{E} t}$. Now, writing the spectral decomposition $Y_\mathrm{E}=\sum_{n}\varepsilon_{n}\ketbra{\varepsilon_{n}}{\varepsilon_{n}}$, we have
\begin{align}
    & \rho_\mathrm{E}^{(ij)}(t) = \sum_{m,n}e^{-i(x_i \varepsilon_{m}-x_j \varepsilon_{n})t}\rho_\mathrm{E}^{m n}\ketbra{\varepsilon_{m}}{\varepsilon_{n}}
\end{align}
where $\rho_\mathrm{E}^{m n}:=\bra{\varepsilon_{n}}\rho_{\mathrm{E},0} \ket{\varepsilon_{m}}$. Examining this together with the total state in Eq.~\eqref{eVonNeuEvolvedState}, we see that the time average of the total state is determined by 
\begin{align}
     g_{mn}^{(ij)}(T):=\left\langle e^{-i(x_i \varepsilon_{m}-x_j \varepsilon_{n})t} \right\rangle_{T} = 
     \begin{cases} 
     1  & \mathrm{if} \;\;  x_i \varepsilon_{m} = x_j \varepsilon_{n}\\
     \dfrac{i e^{[i (x_i \varepsilon_{m}-x_j \varepsilon_{n} )T]}-1}{ (x_i \varepsilon_{m}-x_j \varepsilon_{n} )T} & \mathrm{otherwise}
    \end{cases} 
\end{align}
Assuming ${ x_i \varepsilon_{m} \neq x_j \varepsilon_{n}\;\; \forall i,j,m,n}$ we find an equilibrium state of the desired diagonal form in the pointer basis:
\begin{align}
    \rho_\mathrm{eq} =\sum_i p_i \ketbra{i}{i}_\mathrm{S}\otimes \sum_{m} \rho_\mathrm{E}^{m m}\ketbra{\varepsilon_{m}}{\varepsilon_{m}}.
\end{align}
However, upon examination we see that the environment is independent of $i$, and this state features no correlations between system and environment. The environment therefore can not reveal any information about the system, and such von-Neumann-type Hamiltonians are incapable of (even approximately) equilibrating to states exhibiting objectivity.

\section{Asymptotically approaching Spectrum Broadcast Structure}\label{appsec:typicality}
In this section we show an equilibrium state with Spectrum Broadcast Structure can be obtained approximately by coarse-graining the environment. The purpose of this is to show one example of how Spectrum Broadcast Structure can emerge. This should be considered as a first step; further research is needed to expand this framework into a realistic model of measurements.

As was shown in Section~\ref{sConstrainH}, to obtain a classical-quantum equilibrium state (of which Spectrum Broadcast Structure is an example), one requires a conditional form of the Hamiltonian. In conjunction with the condition of Strong Independence, this prompts us to consider a ``star-shaped" Hamiltonian (see Fig.~1 in the main text) given by 
\begin{equation} \label{eSBSHam}
    H =\sum_{i} \ketbra{i}{i}_{S}\otimes\sum_{k=1}^N c_k H_{k}^{(i)}
\end{equation}
where $H_{k}^{(i)} \in \mathcal{L}\left(\mathcal{H}_k\right)$ and $c_k$ are the different couplings between the system and the observer systems (e.g. representing distance-dependent couplings to spatially distributed observers). For simplicity of notation, we suppress tensor products with the identity operator here, writing $\bigotimes_{j=1}^{k-1} \mathds{1}_j\otimes H_k \otimes\bigotimes_{j'=k+1}^N \mathds{1}_ {j'}$ as simply $H_k$. This generates the unitary evolution
\begin{equation}
    e^{-i H t} = \sum_{i} \ketbra{i}{i}_{S}\bigotimes_{k=1}^{N} U_{k}^{(i)}(t) ,
\end{equation}
where $U_{k}^{(i)}(t):=e^{-i c_k H_{k}^{(i)} t}$.

Assuming a separable initial state, \textit{i.e.} $\rho(0)=\rho_{S,0} \bigotimes_{k=1}^{N} \rho_{k,0}$, the state at time $t$ can be written
\begin{equation}\label{eTimeEvolvedTotal}
    \rho(t) = \sum_{i,j} \bra{i} \rho_{S,0}\ket{j} \ketbra{i}{j}_{S} \bigotimes_{k=1}^{N} \rho_{k}^{(ij)}(t)
\end{equation}
where 
\begin{equation} \label{eTimeDepObservers}
    \rho_{k}^{(ij)}(t) := U_{k}^{(i)}(t) \rho_{k,0} U_{k}^{(j)}(t).
\end{equation}
In order to calculate the equilibrium state $\rho_\mathrm{eq}$, we must therefore calculate the infinite-time average of $\bigotimes_{k=1}^{N} \rho_{k}^{(ij)}(t)$. First, letting $\varepsilon_{n_{k}}^{(i)}$ denote the $n_{k}^\text{th}$ energy eigenvalue of $H_{k}^{(i)}$, we define
\begin{equation}
    \rho_{k}^{mn(ij)}:= \bra{\varepsilon_{m_{k}}^{(i)}} \rho_{k,0} \ket{\varepsilon_{n_{k}}^{(j)}} .
\end{equation}
Then the finite-time average of $\bigotimes_{k=1}^{N} \rho_{k}^{(ij)}(t)$ over the interval $T$ is given by
\begin{gather}
    \left\langle \bigotimes_{k=1}^{N} \rho_{k}^{(ij)}(t) \right\rangle_{T} = \left\langle \bigotimes_{k=1}^{N} \sum_{m_{k}n_{k}} \exp\left[-i c_k \left( \varepsilon_{m_{k}}^{(i)}-  \varepsilon_{n_{k}}^{(j)}\right)t\right] \rho_{k}^{mn(ij)} \ketbra{\varepsilon_{m_{k}}^{(i)}}{\varepsilon_{n_{k}}^{(j)}} \right\rangle_{T}  \nonumber \\
    = \left\langle \left(\sum_{m_{1}n_{1}} \exp\left[-i c_1 \left(  \varepsilon_{m_{1}}^{(i)}-  \varepsilon_{n_{1}}^{(j)}\right)t\right] \rho_{1}^{mn(ij)} \ketbra{\varepsilon_{m_{1}}^{(i)}}{\varepsilon_{n_{1}}^{(j)}} \right) \otimes \left(\sum_{m_{2}n_{2}} \exp\left[-i c_2 \left(  \varepsilon_{m_{2}}^{(i)}- \varepsilon_{n_{2}}^{(j)}\right)t\right] \rho_{2}^{mn(ij)} \ketbra{\varepsilon_{m_{2}}^{(i)}}{\varepsilon_{n_{2}}^{(j)}} \right) \otimes \ldots \right\rangle_{T} \nonumber \\
    = \sum_{m_{1}n_{1}}\sum_{m_{2}n_{2}}\ldots\sum_{m_{N}n_{N}} \left\langle \exp\left[-i\sum_{k=1}^{N} c_k \left( \varepsilon_{m_{k}}^{(i)}- \varepsilon_{n_{k}}^{(j)}\right)t\right] \right\rangle_{T}
    \bigotimes_{k=1}^{N} \rho_{k}^{mn(ij)} \ketbra{\varepsilon_{m_{k}}^{(i)}}{\varepsilon_{n_{k}}^{(j)}} \nonumber \\
    \equiv \sum_{\{m_k\}\{n_k\}} f_{\{m_k\}\{n_k\}}^{(ij)}(T) \bigotimes_{k=1}^{N} \rho_{k}^{mn(ij)} \ketbra{\varepsilon_{m_{k}}^{(i)}}{\varepsilon_{n_{k}}^{(j)}} \label{eTimeAverageEnv}
\end{gather}
where we have used the notation
\begin{align}
    \sum_{\{a_k\}}=\sum_{a_1}\sum_{a_2}...\sum_{a_N},
\end{align}
and defined
\begin{equation} \label{eTimeAverageFunction}
    f_{\{m_k\}\{n_k\}}^{(ij)} (T) :=\begin{cases} 1 &\text{if }\, \sum_{k=1}^{N} c_k \varepsilon_{m_{k}}^{(i)}= \sum_{k=1}^{N} c_k \varepsilon_{n_{k}}^{(j)} \\
    i \frac{\exp\left[-i\sum_{k=1}^{N}c_k \left(  \varepsilon_{m_{k}}^{(i)}- \varepsilon_{n_{k}}^{(j)}\right)T\right]-1}{\sum_{k=1}^{N}c_k \left(  \varepsilon_{m_{k}}^{(i)}- \varepsilon_{n_{k}}^{(j)}\right)T} & \text{otherwise.} \end{cases}
\end{equation}
Noting that $\lbrace\sum_{k=1}^{N} c_k \varepsilon_{m_{k}}^{(i)}\rbrace_{i,m_{1},\ldots,m_{N}}$ are the eigenvalues of $H$, which can be seen from Eq.~\eqref{eSBSHam}, we see that the cases where $f^{(ij)}_{\{m_k\}\{n_k\}}=1$ for $i\neq j$ and $\{m_k\}\neq \{n_k\}$ correspond to degeneracies in $H$. Further note that such cases are particularly ``fine-tuned'', as the coupling constants and energy eigenvalues have to conspire in order to yield this result. Thus, assuming that the total Hamiltonian is non-degenerate, the first case in Eq.~\eqref{eTimeAverageFunction} never occurs, and we have
\begin{equation}
    \left\langle \bigotimes_{k=1}^{N} \rho_{k}^{(ij)}(t) \right\rangle_{\infty} = 0 \qquad \text{ for } i \neq j .
\end{equation}
Combining this with Equations~\eqref{eTimeEvolvedTotal} and~\eqref{eTimeAverageEnv}, we obtain
\begin{gather} \label{eq:InfTimeDiag} 
    \left\langle \rho(t) \right\rangle_{\infty} = \sum_{i} \bra{i} \rho_{S,0}\ket{i} \ketbra{i}{i}_{S} \left\langle \bigotimes_{k=1}^{N} \rho_{k}^{(ii)}(t) \right\rangle_{\infty} \nonumber \\
    = \sum_{i} p_{i} \ketbra{i}{i}_{S} \sum_{\{m_k\}\{n_k\}} f_{\{m_k\}\{n_k\}}^{(ii)} (\infty) \bigotimes_{k=1}^{N} \rho_{k,0}^{mn(ii)} \ketbra{\varepsilon_{m_{k}}^{(i)}}{\varepsilon_{n_{k}}^{(i)}},
\end{gather}
where $f^{(ii)}_{\{m_k\}\{n_k\}}(\infty):=\lim_{T\rightarrow\infty} f^{(ii)}_{\{m_k\}\{n_k\}}(T)$, and again $p_i :=  \bra{i} \rho_{S,0}\ket{i}$ is the probability of outcome $i$ for the projective measurement $\lbrace\ketbra{i}{i}_{S} \rbrace_{i}$ on the system. Due to the assumption of non-degeneracy of the total Hamiltonian, the conditional Hamiltonians $H_k^{(i)}$ have to be non-degenerate as well, and consequently, $\sum_{k=1}^{N} c_k \varepsilon_{m_{k}}^{(i)} \neq \sum_{k=1}^{N} c_k \varepsilon_{n_{k}}^{(i)}$ $\forall \, m_{k}\neq n_{k}$, and thus only the $m_{k}= n_{k}$ terms in $f^{(ii)}_{\{m_k\}\{n_k\}}(\infty)$ are nonzero. Equation~\eqref{eTimeAverageFunction} then implies that
\begin{gather}
    \rho_\mathrm{eq} = \sum_{i} p_{i} \ketbra{i}{i}_{S} \sum_{m_{1}\ldots m_{N}} \bigotimes_{k=1}^{N} \rho_{k}^{mm(ii)} \ketbra{\varepsilon_{m_{k}}^{(i)}}{\varepsilon_{m_{k}}^{(i)}} \nonumber \\
    = \sum_{i} p_{i} \ketbra{i}{i}_{S}  \bigotimes_{k=1}^{N} \sum_{m_{k}} \rho_{k}^{mm(ii)} \ketbra{\varepsilon_{m_{k}}^{(i)}}{\varepsilon_{m_{k}}^{(i)}} \nonumber \\
    \equiv \sum_{i} p_{i} \ketbra{i}{i}_{S}  \bigotimes_{k=1}^{N} \rho_{k}^{(i)},
\end{gather}
where 
\begin{equation}
    \rho_{k}^{(i)} := \sum_{m_{k}} \ketbra{\varepsilon_{m_{k}}^{(i)}}{\varepsilon_{m_{k}}^{(i)}} \, \rho_{k,0} \, \ketbra{\varepsilon_{m_{k}}^{(i)}}{\varepsilon_{m_{k}}^{(i)}}
\end{equation}
is the initial state of observer system $k$ (\textit{i.e.} $\rho_{k,0}$) subject to the pinching map with respect to $H_{k}^{(i)}$. Now let us gather together the $N$ factors of the environment into $M$ ``macro-observers'', and let $N_{q}$ denote the set of environment labels belonging to the $q^\text{th}$ macro-observer, so that ${\bigotimes_{k=1}^{N}\mathcal{H}_k=\bigotimes_{q=1}^{M}\mathcal{E}_q}$ with $\mathcal{E}_{q}:=\bigotimes_{k \in N_{q}}\mathcal{H}_k$. Then
\begin{gather} \label{ePreSBSstate}
    \rho_\mathrm{eq}  = \sum_{i} p_{i} \ketbra{i}{i}_{S}  \bigotimes_{q=1}^{M} \tilde{\rho}_{q}^{(i)}, \qquad \text{ where } \qquad \tilde{\rho}_{q}^{(i)} := \bigotimes_{k \in N_{q}} \rho_{k}^{(i)} ,
\end{gather}
We now show that the fidelity $F \left( \tilde{\rho}_q^{(i)},\tilde{\rho}_q^{(j)} \right)$ between conditional states $\tilde{\rho}_{q}^{(i)}$ and $\tilde{\rho}_{q}^{(j)}$ of each macro-observer $q$ tends rapidly to 0 as the macro-observer size increases. To do so, we use the inequality
\begin{align}
    \sqrt{F(\sigma,\rho)}\leq \sum_m \sqrt{\tr[E_m\rho]}\sqrt{\tr[E_m\sigma]}
\end{align}
where the $\lbrace E_{m}\rbrace$ are the  elements of an arbitrary POVM, to bound $F \left( \rho_k^{(i)},\rho_k^{(j)} \right)$. Choosing the POVM elements to be projectors onto the eigenbasis of $H_{k}^{(i)}$, \emph{i.e.} $\lbrace \ketbra{\varepsilon_{m_k}^{(i)}}{\varepsilon_{m_k}^{(i)}}\rbrace_{m_{k}}$. Since $\rho_{k}^{(i)}=\sum_{m_k} \rho_{k}^{mm(ii)}\ketbra{\varepsilon_{m_k}^{(i)}}{\varepsilon_{m_k}^{(i)}}$ we obtain
\begin{align} \label{eMicroFidelityInequ}
   & \sqrt{F(\rho_{k}^{(i)},\rho_{k}^{(j)})}\leq \sum_{m_k}\sqrt{\rho_{k}^{mm(ii)}}\sqrt{\tr[\sum_{n_k}\rho_{k}^{nn(jj)}\ketbra{\varepsilon_{n_k}^{(j)}}{\varepsilon_{n_k}^{(j)}}\varepsilon_{m_k}^{(i)}\rangle\!\langle \varepsilon_{m_k}^{(i)}|]}\nonumber\\&
     =\sum_{m_k}\sqrt{\rho_{k}^{mm(ii)}}\sqrt{\sum_{n_k}\rho_{k}^{nn(jj)}|\braket{\varepsilon_{m_k}^{(i)}}{\varepsilon_{n_k}^{(j)}}|^2}=\sum_{m_k}\sqrt{\sum_{n_k}\rho_{k}^{mm(ii)}\rho_{k}^{nn(jj)}|\braket{\varepsilon_{m_k}^{(i)}}{\varepsilon_{n_k}^{(j)}}|^2} \nonumber \\
     & = \sum_{m_k}\sqrt{\bra{\varepsilon_{m_k}^{(i)}} \rho_{k}^{(i)}\rho_{k}^{(j)} \ket{\varepsilon_{m_k}^{(i)}}}.
\end{align}
In the cases where $H^{(i)}=H^{(j)}$ or $\rho_{k,0}=\dfrac{1}{d_k}\mathds{1}$ we see that the fidelity is bounded above by 1, as one would expect. 

To see that $F \left( \tilde{\rho}_q^{(i)},\tilde{\rho}_q^{(j)} \right)$ approaches zero exponentially as the macro-observers grow, we recall that $F(\bigotimes_k\sigma_k,\bigotimes_k\rho_k)=\prod_k F(\sigma,\rho)$. Combining this with the inequality in~\eqref{eMicroFidelityInequ}, we have
\begin{align}
    \sqrt{F(\bigotimes_{k\in N_q}\rho_{k}^{(i)},\bigotimes_{k\in N_q}\rho_{k}^{(j)})}&=\prod_{k\in N_q}\sqrt{F(\rho_{k}^{(i)},\rho_{k}^{(j)})} \nonumber \\
    &\leq \prod_{k\in N_q}\left(\sum_{m_k}\sqrt{\bra{\varepsilon_{m_k}^{(i)}} \rho_{k}^{(i)}\rho_{k}^{(j)} \ket{\varepsilon_{m_k}^{(i)}}}\right) \nonumber \\
    & \equiv \prod_{k\in N_q} \eta_k^{(ij)} ,
\end{align}
where we have defined $\eta_k^{(ij)}:=\sum_{m_k}\sqrt{\bra{\varepsilon_{m_k}^{(i)}} \rho_{k}^{(i)}\rho_{k}^{(j)} \ket{\varepsilon_{m_k}^{(i)}}}$. Further defining $N_q^{(ij)}:=\{k\in N_q^{(ij)}:\eta_k^{(ij)}<1\}$ and $\eta_{max}^{(ij)}:=\max_{k\in N_q^{(ij)}}\eta_k^{(ij)}$. We then have
\begin{align}
    F(\tilde{\rho}_{q}^{(i)},\tilde{\rho}_{q}^{(j)})\leq \left(\eta_{max}^{(ij)}\right)^{2|N_q^{(ij)}|},
\end{align}
which can be simplified by defining $\gamma^{(ij)}_q=-2 \ln(\eta_\mathrm{max}^{(ij)})$ to finally obtain
\begin{align} \label{eMacroFidelityInequ}
     F(\tilde{\rho}_{q}^{(i)},\tilde{\rho}_{q}^{(j)}) \leq e^{-\gamma^{(ij)}_q |N_q^{(ij)}| } .
\end{align}
Assuming that the conditional Hamiltonians $H_k^{(i)}$ and $H_k^{(j)}$ have at least one distinct eigenvector, the upper bound in~\eqref{eMacroFidelityInequ} exponentially tends to 0 for increasing macro-observer size. The equilibrium state, therefore, tends exponentially to one with Spectrum Broadcast Structure state as the size of the macro-observers (and thus the environment as a whole) increases. Further assuming that $H_k^{(i)}$ and $H_k^{(j)}$ have no eigenvectors in common, we find the simpler result stated in the main text, \textit{i.e.}
\begin{align}
     F(\tilde{\rho}_{q}^{(i)},\tilde{\rho}_{q}^{(j)}) \leq e^{-\gamma^{(ij)}_q |N_q| } ,
\end{align}
where $|N_q|$ is the number of observer systems $k$ collected into the macro-observer $q$.


\bibliographystyle{apsrev4-1fixed_with_article_titles_full_names.bst}
\bibliography{bibfile.bib}
\end{document}